\documentclass[12pt]{article}
\usepackage{amsmath}
\usepackage{graphicx}
\usepackage{natbib}
\usepackage{url} 
\usepackage{amssymb}
\usepackage{color}
\usepackage{xcolor}
\usepackage{bm}
\usepackage[unicode,colorlinks=true,pagebackref=false]{hyperref}
\usepackage{listings}
\usepackage[normalem]{ulem}

\usepackage{apptools}
\AtAppendix{\counterwithin{figure}{section}}
\AtAppendix{\counterwithin{lstlisting}{section}}

\usepackage{mdframed} 
\definecolor{light-gray}{gray}{0.95} 

\lstdefinestyle{myListingStyle} 
    {
        basicstyle = \linespread{1}\small\ttfamily,
        backgroundcolor = \color{light-gray},
    }

\newcommand\ale[1]{\textcolor{black}{#1}}

\newcommand{\blind}{0}

\begin{document}

\def\spacingset#1{\renewcommand{\baselinestretch}%
{#1}\small\normalsize} \spacingset{1}


\if0\blind
{
  \title{\bf Probabilistic Programming with Sufficient Statistics for faster Bayesian Computation}
  \author{Clemens Pichler
    \\
    Institute of Statistics and Mathematical Methods in Economics,\\ 
    TU Wien,  Vienna, Austria\\
    and \\
    Jack Jewson\thanks{Equal contribution} \\
    Department of Econometrics and Business Statistics,\\
    Monash University, Melbourne, Australia\\
    and \\
    Alejandra Avalos-Pacheco${}^{*}$ \\
    Institute of Applied Statistics, JKU Linz, Linz, Austria,\\
    Harvard-MIT Center for Regulatory Science, Harvard University,\\
    Boston, MA, USA
    }
  \maketitle
} \fi

\if1\blind
{
  \bigskip
  \bigskip
  \bigskip
  \begin{center}
    {\LARGE\bf Probabilistic Programming with Sufficient Statistics for faster Bayesian Computation}
\end{center}
  \medskip
} \fi

\bigskip
\begin{abstract}
Probabilistic programming methods have \ale{revolutionised} Bayesian inference, \ale{making it easier than ever for practitioners to perform } Markov-chain-Monte-Carlo sampling from non-conjugate posterior distributions. Here we focus on \texttt{Stan}\ale{, arguably the most used probabilistic programming tool for Bayesian inference} \citep{carpenter2017stan}, and its interface with \textit{R} via the \texttt{brms} \citep{burkner2017brms} and \texttt{rstanarm} \citep{goodrich2024rstanarm} packages. Although easy to implement, \ale{
these tools can become computationally prohibitive when applied to datasets with many observations or models with numerous parameters.}

\ale{While the use of sufficient statistics is well-established in theory, it has been surprisingly overlooked in state-of-the-art \texttt{Stan} software.
We show that when the likelihood can be written in terms of sufficient statistics, considerable computational improvements can be made to current implementations.
} 
We demonstrate \ale{how this approach provides accurate inference at a fraction of the time than state-of-the-art implementations} for Gaussian linear regression models with non-conjugate priors, hierarchical random effects models, and factor analysis models.  
\ale{Our results also show that moderate computational gains can be achieved even in models where the likelihood can only be partially written in terms of sufficient statistics.}
\end{abstract}

\noindent%
{\it Keywords:} 
Exponential Families, Markov-chain-Monte-Carlo, \texttt{Stan}, \texttt{brms}, \texttt{rstanarm}
\vfill

\newpage
\spacingset{1.75} 
\section{Introduction}
\label{Sec:Introduction}

Bayesian inference has become an important tool in modern statistics, providing a robust framework for updating beliefs in the presence of new data.
Advantages of the Bayesian approach include: 
(i) arguably more flexible tools than frequentist 
methods \citep{Berry2010BAMf}, (ii) useful procedures to obtain posterior predictive probabilities 
that can be updated as new data become available \citep{Berger1988SAat}, (iii) 
natural ways to incorporate expert knowledge, by using informative priors and hierarchical models \citep{Berry2010BAMf}, (iv) a rigorous, realistic, and easy-to-interpret decision-theoretic framework \citep{Berger1985Sdta}, 
and (v) uncertainty quantification of the parameters of interest through the posterior \citep{Berry2010BAMf}.

However, the application of Bayesian inference is often limited by the computational effort required to obtain the posterior distribution. Beyond simple conjugate models, the posterior distribution is not available in closed form and must therefore be approximated. Markov-chain-Monte-Carlo (MCMC) methods \citep{gelfand1990sampling}, which construct a Markov chain whose stationary distribution is the target posterior, have been extensively used to generate posterior samples and approximate posterior expectations. However, these algorithms often require carefully designed proposal distributions and fine-tuning, which can be a significant obstacle for non-expert users.

Probabilistic programming languages 
mitigate these challenges by abstracting away much of the complexity. They allow users to specify their prior and likelihood functions directly, similar to how one might write a Bayesian model on a whiteboard (see e.g. Listing \ref{Lst:stan_gaussian}). The model is then compiled, and a MCMC algorithm is automatically tuned to sample from the model's posterior distribution. 
Leading examples of probabilistic programming languages are \texttt{Stan} \citep{carpenter2017stan}, \texttt{PyMC} \citep{abril2023pymc}, \texttt{Numpyro} \citep{phan:2019}, and \texttt{nimble} \citep{de2017programming}.
We focus specifically on the \texttt{Stan} probabilistic programming language \citep{carpenter2017stan}.

\ale{\texttt{Stan} is arguably the  most widely used tool in academic research and industry, often considered the gold standard for Bayesian modeling. 
Currently it has more than 100,000 users, and it interfaces with
\textit{R}, \textit{Python}, \textit{Julia}, \textit{Scala}, \textit{Stata}, \textit{Matlab}, command line interfaces, as well as many packages to support diagnostics and workflow.
\texttt{Stan} employs state-of-the-art inference algorithms, including Hamiltonian Monte Carlo (HMC) and its variants, providing efficient and accurate sampling for complex models.
Here, we focus on \texttt{Stan}’s integration} with \textit{R} via \texttt{Rstan} and \ale{packages such as} the \texttt{brms} and \texttt{rstanarm} which further simplify implementation by automatically generating `optimized' \texttt{Stan} files for popular Bayesian models. 

Although 
\ale{\texttt{Stan}} has significantly improved the accessibility of Bayesian inference, as models grow in complexity and dimensionality, it can be computationally burdensome, and in some cases prohibitive. In scenarios with a large number of observations, $n$, a significant computational bottleneck in Bayesian inference is the evaluation of the log-likelihood. However, many popular statistical models originate from the exponential family and therefore omit a representation in terms of sufficient statistics. The model's sufficient statistics are only functions of the observed data and can be precomputed, this allows the likelihood to be evaluated as one function of the sufficient statistics and parameters and avoids the costly sum of  $n$ terms.

While this simplification is well-known, it has been overlooked in the context of \texttt{Stan}. For example, in Bayesian linear regression under a non-conjugate prior, the implementations provided in the Stan User Guide 
\citep{stan2024stan}, \texttt{brms} package \citep{burkner2017brms} and \texttt{rstanarm} package \citep{goodrich2024rstanarm} all fail to take advantage of this simplification and are therefore unnecessarily computationally demanding.

\ale{We show that when the likelihood can be fully expressed in terms of sufficient statistics, considerable improvements in computational efficiency can be achieved compared to current implementations. 
Furthermore, our approach delivers accurate inference (see Figures \ref{Fig:regression_posteriors}, \ref{Fig:mixed_posteriors}, \ref{Fig:factor_posteriors}, \ref{Fig:poisson_posteriors}) at a fraction of the time (see Figures \ref{Fig:regression}, \ref{Fig:random_effects}, \ref{Fig:factor_analysis}) required by state-of-the-art methods, as demonstrated in Gaussian linear regression models with non-conjugate priors, hierarchical random effects models, and factor analysis models. 
Notably, we also find that moderate computational gains are attainable even in scenarios where the likelihood can only be partially written in terms of sufficient statistics (see Figure \ref{Fig:poisson_regression}), highlighting the broad applicability and robustness of this methodology.}

The rest of this article is organized as follows: Section \ref{Sec:methods} reviews Bayesian inference, exponential family models, sufficient statistics and probabilistic programming languages, Section \ref{Sec:Examples} provides three concrete examples where writing the likelihood in terms of sufficient statistics provides considerable speed up to default applications. We additionally provide an example where the likelihood can only partially be written in terms of  sufficient statistics and small computational gains can still be achieved. Section \ref{Sec:discussion} concludes.
Code to reproduce our experiments is available in the supplementary material and at 
\url{https://github.com/jejewson/ProbabilisticProgrammingSufficientStatistics}

\section{Bayesian Inference, Exponential Families and Probabilistic Programming}{\label{Sec:methods}}

\subsection{Exponential Families and Sufficient Statistics}

Consider observations $\bm{y} = (\bm{y}_1, \ldots, \bm{y}_n) \in \mathcal{Y}^{n\times d} \subseteq \mathbb{R}^{n\times p}$, assumed to be independently and identically distributed according to \{$F(\cdot; \bm{\theta})$, $\bm{\theta}\in\Theta\subseteq\mathbb{R}^p$\}. \{$F(\cdot; \bm{\theta})$, $\bm{\theta}\in\Theta\subseteq\mathbb{R}^p$\} is an exponential family distribution \citep{pitman1936sufficient, darmois1935lois, koopman1936distributions} if its density or mass function can be written as 
\begin{equation}
f(\bm{y} ; \bm{\theta}) = h(\bm{y}) \exp \left( \eta(\bm{\theta})^\top T(\bm{y}) - A(\bm{\theta}) \right)\nonumber
\end{equation}
where $\eta: \mathbb{R}^p\mapsto \mathbb{R}^q$ maps the parameters to the natural parameters,  $T: \mathbb{R}^d\mapsto \mathbb{R}^q$ maps each observation to its sufficient statistics, $A:\mathbb{R}^p\mapsto \mathbb{R}$ ensures normalisation to 1, and $h:\mathbb{R}^d\mapsto\mathbb{R}_{+}$. The Gaussian, Poisson, gamma, beta and binomial distributions are members of the exponential family. 

A convenient feature of exponential family distributions is that the resulting log-likelihood function for $n$ independent and identically distributed (\textit{iid}) observations $\bm{y}$ simplifies to,
\begin{equation}
\ell(\bm{y};  \bm{\theta}) := \sum_{i=1}^n \log f(\bm{y}_i ; \bm{\theta}) = \sum_{i=1}^n\{\log(h(\bm{y}_i))\} + \eta(\bm{\theta})^\top s(\bm{\theta}) - nA(\bm{\theta}).\nonumber
\end{equation}
with $s(\bm{y}) := \sum_{i=1}^n \{T(\bm{y}_i)\}$. Therefore, precomputing $s(\bm{y})$ means that evaluating $\ell(\bm{y};  \bm{\theta})$ ignoring terms that do not depend on $\bm{\theta}$ can be done independently of $n$, i.e. the same computation is required when $n= 10$ as when $n = 10^6$.

\subsection{Bayesian Inference}{\label{Sec:Bayes}}

Placing prior $\pi(\bm{\theta})$ on the model parameter $\bm{\theta} \in \Theta$, the posterior distribution for $\bm{\theta}$ after observing $\bm{y}$ is given by:
\begin{equation}
    \pi(\bm{\theta} \mid \bm{y}) = \frac{\pi(\bm{\theta})\prod_{i=1}^n f(y_i;\bm{\theta})}{\int \pi(\bm{\theta})\prod_{i=1}^n f(y_i;\bm{\theta}) d\bm{\theta}} = \frac{\pi(\bm{\theta})\exp\left\{\sum_{i=1}^n \ell(y_i;\bm{\theta})\right\}}{\int \pi(\bm{\theta})\exp\left\{\sum_{i=1}^n \ell(y_i;\bm{\theta})\right\} d\bm{\theta}}.\label{Equ:BayesRule}
\end{equation}
The posterior $\pi(\bm{\theta} \mid \bm{y})$ is used to calculate posterior expected values, high-density sets for parameters, and predictive distributions for future observations.
The tractability of $\pi(\bm{\theta} \mid \bm{y})$ depends on whether the normalising constant $\int \pi(\bm{\theta})\exp\left\{\sum_{i=1}^n \ell(y_i;\bm{\theta})\right\} d\bm{\theta}$ can be computed. An advantage of exponential family models is the existence of a conjugate prior \citep{diaconis1979conjugate} which leaves the posterior in the same family as the prior and facilitates straightforward computation. 

However, there exist many examples where even though the likelihood is an exponential family, additional structure such as latent variables i.e. in random effects or factor models, mean that no fully conjugate prior exists. Further, choosing a prior solely for computationally convenient reasons is at odds with the foundation of Bayesian analyses \citep{goldstein2006subjective} and their rigid form may restrict the ability to encode reasonable prior beliefs. For example, non-conjugate heavy tailed weakly-informative Cauchy and Student-$t$ priors \citep{gelman2013bayesian, gelman2008default} have become default priors for regression models.


When a conjugate prior is not available, or where conjugate priors do not sufficiently capture the complexities of the prior information, MCMC \citep{gelfand1990sampling} is the standard method to approximate the posterior. MCMC methods sample from a Markov chains whose stationary distribution is the posterior, and then use these samples to approximate posterior expectations via the law of large numbers. The Metropolis-Hastings algorithm \citep{metropolis1953equation, hastings1970monte}, in particular, allows this to be done without requiring the computation of the intractable normalising constant in \eqref{Equ:BayesRule}.

Such methods, however, make Bayesian inference computationally demanding. Approximating the posterior for even moderate dimensional $\bm{\theta}$ requires thousands of samples, and the Markov chain may need to run for many more iterations to obtain these. A key bottleneck of most MCMC algorithms is that $\ell(\bm{y}; \bm{\theta})$ must be evaluated at least once per iteration e.g. when computing the Metropolis-Hastings acceptance ratio. As a result, even when non-conjugate priors are employed, writing the likelihood in terms of sufficient statistics, where possible, continues to provide computational savings.

\subsection{Probabilistic Programming}

Initially, obtaining samples that well approximate the target posterior in a reasonable amount of time required careful selection, implementation and tuning of the MCMC algorithm, from a plethora of available. This presented a significant barrier for applied practitioners wishing to undertake Bayesian analyses. Probabilistic programming languages, e.g. \texttt{Stan}  \citep{carpenter2017stan}, \texttt{PyMC} \citep{abril2023pymc}, \texttt{Numpyro} \citep{phan:2019}, and \texttt{nimble} \citep{de2017programming}, have removed this barrier. Probabilistic programming languages allow the user to write out their Bayesian model in a straightforward format, similar to how it would be written on a whiteboard, and then they automatically tune a MCMC sampler to sample from the implied posterior.

We focus specifically on \texttt{Stan}. 
\texttt{Stan} primarily uses the No-U-Turn Sampler (NUTS) \citep{hoffman2014no}, an adaptive form of Hamiltonian Monte Carlo (HMC) \citep{duane1987hybrid}, for posterior sampling. HMC introduces an auxiliary ``momentum" variable and leverages the principles of Hamiltonian dynamics to generate proposals that can move through high dimensional parameter space more effectively than traditional random walk approaches, see \cite{neal2011mcmc}. The No-U-Turn Sampler (NUTS) \citep{hoffman2014no} automatically tunes the hyperparameters of HMC to avoid ``U-turning'' trajectories and wasteful computation.



Listing \ref{Lst:stan_gaussian} presents \texttt{Stan} code for a Gaussian location model with non-conjugate Cauchy prior and closely resembles how one would write out such a model on a whiteboard. 
The \textit{R} package \texttt{RStan} \citep{guo2020package} allows users to compile their \texttt{Stan} models and run them in \textit{R}. \textit{R} packages such as \texttt{brms} \citep{burkner2017brms} and \texttt{rstanarm} \citep{goodrich2024rstanarm} simplify to use of \texttt{Stan} even further by running `optimized' default \texttt{Stan} files for popular models. A differentiator between the two is that \texttt{brms} generates the \texttt{Stan} code for the user while \texttt{rstanarm} runs this in the background via \texttt{cmdrstan}. Other interfaces with \texttt{Stan} include the \texttt{rethinking} package \citep{mcelreath2018statistical}.

While contributing immensely to the usability of Bayesian inference, it appears as though the packages mentioned previously fail to take advantage of the computational advancements introduced in Section \ref{Sec:Bayes} for several popular models. Thus in their current form, their ease of use comes at an unnecessary computational cost to the user.

\section{Faster Probabilistic Programming}{\label{Sec:Examples}}

We demonstrate three popular models where taking advantage of sufficient statistics leads to considerable time savings over the current implementations. A further example is provided showing that even when the likelihood cannot be completely written in terms of sufficient statistics computational savings can be made.

\subsection{Gaussian Linear Regression}{\label{Sec:Regression}}

Gaussian linear regression posits that the conditional density of univariate observations $y_i\in\mathbb{R}$ given $p$-dimensional predictors $\bm{x}_i\in\mathbb{R}^p$ (that may include a constant 1 for the intercept) is $y_i\mid \bm{x}_i \sim\mathcal{N}\left(\bm{x}_i^\top\bm{\beta}, \sigma^2\right)$ where $\bm{\beta}\in\mathbb{R}^p$ are the regression coefficients and $\sigma^2$ is the residual variance. 
The full parameter vector is $\bm{\theta} = \{\bm{\beta}, \sigma^2\}$. This is an exponential family model and its log-likelihood can be written as
\[
\ell(\bm{y}; \bm{\beta}, \sigma, X) = -\frac{n}{2} \log(\sigma) - \frac{1}{2\sigma^2} (S_{yy} - 2\bm{\beta}^\top S_{yx} + \bm{\beta}^\top S_{xx}\bm{\beta})
\]
where $X\in\mathbb{R}^{n\times p}$ is a matrix whose $i$th row is $\bm{x}_i$ and  \( S_{xx} = X^\top X \), \( S_{yy} = \bm{y}^\top\bm{y} \), and \( S_{yx} = \bm{y}^\top X \) are the sufficient statistics.

Although the normal-inverse-gamma prior is conjugate here, heavier-tailed Half-Cauchy or Half-Student-$t$ priors are often preferred to provide weakly informative priors for the residual variance. As a result, MCMC is required to sample from the induced posterior.

We compare a vectorised implementation from the Stan User Guide (Listing \ref{Lst:stan_regression_userguide}), the \texttt{Stan} code produced by \texttt{brms} (Listing \ref{Lst:stan_regression_brms}) and \texttt{rstanarm}'s implementation of linear regression  with an implementation that takes advantage of sufficient statistics (Listing \ref{Lst:stan_regression_suffstat}). For all models, we adopt the default prior implemented in \texttt{brms} $\sigma \sim t_{3}(0, 3.7)$ where $t_{\nu}(\mu, s)$ is a Student's-$t$ density with degrees of freedom $\nu$, location $\mu$ and scale $s$, and set $\beta_j \sim \mathcal{N}(0, 10^2)$, $j=1,\ldots, p$ a priori. Figure \ref{Fig:regression_posteriors} overlays the posterior approximations generated by the different \texttt{Stan} implementations and confirms that all are sampling from the same posterior.

We generated $n \in \{100, 1000, 10000\}$ observations for a linear regression model $y_i = \bm{\beta}^TX_i + \epsilon$ with $p\in \{10, 100, 500\}$ dimensional predictors simulated such that $X_{ij}\sim \mathcal{N}(0, 1)$, regression coefficients $\beta_1 = 1.5$, $\beta_2 = 2$, $\beta_3 = 2.5$ and $\beta_j = 0$, $j = 4, \ldots, p$ and  $\epsilon \sim \mathcal{N}(0, 1)$.

\begin{figure}[!ht]
\begin{center}
\includegraphics[trim= {0.0cm 0.00cm 0.0cm 0.0cm}, clip,  width=0.99\columnwidth]{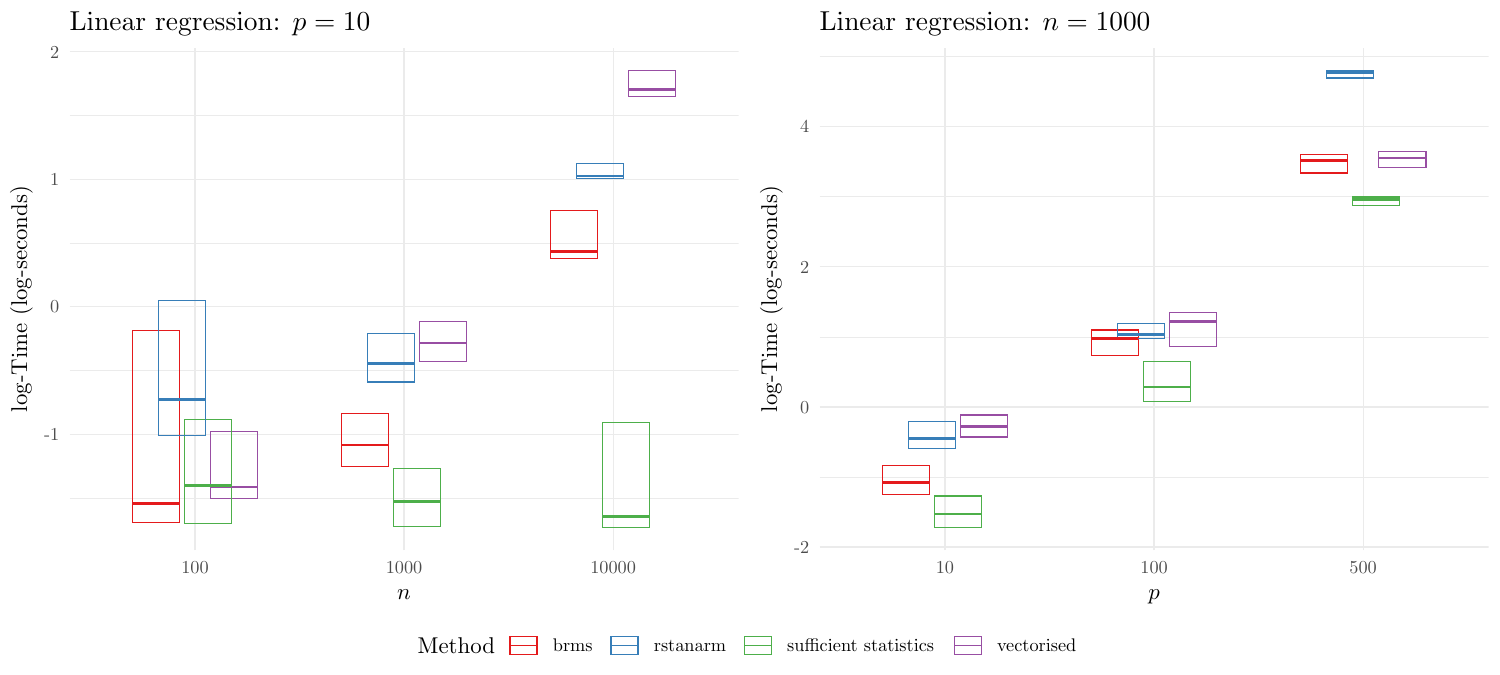}
\caption{Time taken to produce 5000 posterior samples after 1000 warm-up iterations from a Bayesian linear regression model with non-conjugate priors using different \texttt{Stan} implementations. \textbf{Left}: Increasing number of observations $n$. \textbf{Right}: Increasing number of regression parameters $p$.}
\label{Fig:regression}
\end{center}
\end{figure}

Figure \ref{Fig:regression} compares the time taken to produce 5000 post warm-up samples after a 1000 sample warm-up period for the different implementations repeated 25 times using the \texttt{microbench} package \citep{mersmann2024microbench}. As the number of observations $n$ increases the time taken by our sufficient statistics implementation remains constant, as expected, while the time required by the other methods increases considerably. As $p$ increases, the time taken for all four methods increases, but the sufficient statistics version remains the fastest.


\subsection{Mixed Effects Models}{\label{Sec:RandomEffects}}



Hierarchical models such as mixed linear effects models \citep[see e.g.][]{oberg2007linear, gelman2006data} extend regression models to account for both fixed and random variability across different groups or clusters in hierarchical or nested data structures. These models are particularly useful when observations within the same group are more similar to each other than to those in other groups, while also allowing for fixed effects that capture relationships at the population level. A common example of this is when analysing test results across multiple students in different schools \citep{gelman2006data}. 

Consider a dataset with $J$ groups with each group containing $n_j$ observations. Let \(y_{ij}\) represent the \(i\)-th observation in group \(j\), and let \(\bm{x}_{ij}\) be the corresponding $p$-dimensional (with first element 1 if the model has an intercept) predictor variable. An example of a mixed effects model is
\begin{equation}
y_{ij} \sim\mathcal{N}\left( \bm{x}_{ij}^\top\bm{\beta} + u_j, \sigma^2\right), \quad u_j \sim \mathcal{N}(0, \sigma_u^2)\nonumber
\end{equation}
where  \(\beta\in\mathbb{R}^p\) are the regression coefficients (fixed effects), \(u_j\) is the random effect associated with group \(j\), $\sigma^2$ is the observation-level residual variance, and \(\sigma_u^2\) is the variance of the random effects across groups. The full parameter vector is $\bm{\theta} = \{\bm{\beta}, \bm{u}, \sigma^2, \sigma^2_u\}$. The hierarchical nature of this model means there is no conjugate prior, however, the likelihood is still from an exponential family and, therefore, we can write
\begin{equation}
\begin{aligned}
&\ell(\bm{y}; \bm{\beta}, \sigma, \sigma_u, \bm{u}) = -\frac{J}{2} \log (2\pi\sigma_u^2) - \frac{\bm{u}^T\bm{u}}{2\sigma_u^2}  - \frac{\bm{n}^\top\bm{1}}{2} \log (2\pi\sigma^2)\nonumber\\
&- \frac{1}{2\sigma^2} \Bigg\{ \left(S_{yy} + (\bm{u}\odot\bm{u})^\top \bm{n} - 2 \bm{u}^T \bar{\bm{y}} \right) - 2 \left(S_{yx} - \bm{u}^T\bar{X} \right) \bm{\beta} + \bm{\beta}^\top S_{xx} \bm{\beta} \Bigg\},
\end{aligned}
\end{equation}
where $\odot$ is an element-wise multiplication and $S_{yy} := \sum_{j=1}^J \sum_{i=1}^{n_j} y_{ij}^2 = \bm{y}^T\bm{y}$, where $\bm{y}\in\mathbb{R}^{\sum_{j=1}^Jn_j}$ is a vector of the stacked $y_{ij}$'s, $S_{yx} := \sum_{j=1}^J \sum_{i=1}^{n_j} \bm{x}_{ij} y_{ij} = \bm{y}^TX$, where $X\in \mathbb{R}^{\sum_{j=1}^Jn_j \times p}$ is a matrix of the stacked $\bm{x}_{ij}$'s, $S_{xx} := \sum_{j=1}^J \sum_{i=1}^{n_j} \bm{x}_{ij}\bm{x}_{ij}^\top = X^TX$, 
$\bar{\bm{y}} = (\bar{y}_1, \ldots, \bar{y}_J)^T$ with 
$\bar{y}_j := \sum_{i=1}^{n_j} y_{ij}$, $\bar{X} \in \mathbb{R}^{J\times p}$ has rows $\bar{\bm{x}}_j := \sum_{i=1}^{n_j} \bm{x}_{ij}$, $j = 1,\ldots, J$, and $\bm{n} = (n_1, \ldots, n_J)$ are the sufficient statistics.

We compare a vectorised implementation (Listing \ref{Lst:stan_hierarchical_vectorised}), the \texttt{Stan} code produced by \texttt{brms} (Listing \ref{Lst:stan_hierarchical_brms}) and \texttt{rstanarm}'s implementation with an implementation that takes advantage of sufficient statistics (Listing \ref{Lst:stan_hierarchical_SuffStat}). 
For all models, we adopt the default prior implemented in \texttt{brms} $\sigma, \sigma_u \sim t_{3}(0, 3.7)$, and set $\beta_j \sim \mathcal{N}(0, 10^2)$, $j=1,\ldots, p$ a prior. Note, that \texttt{rstanarm} has limited flexibility when specifying the prior for $\sigma_u$, and as a result this prior is not identical to that used in the other methods. Figure \ref{Fig:mixed_posteriors} overlays the posterior approximations generated by the different \texttt{Stan} implementations and confirms that all are sampling from the same posterior with the exception of $\sigma_u$ in the \texttt{rstanarm} implementation.

We generated $n \in \{100, 1000, 10000\}$ observations, allocated uniformly at random to $J \in \{5, 50, 250\}$ groups from a linear mixed effects model $y_{ij} = \bm{\beta}^TX_{ij} + u_j + \epsilon$ with $p = 5$ dimensional predictors simulated such that $X_{ij}\sim \mathcal{N}(0, 1)$, regression coefficients $\bm{\beta} = (1.5, 2, 2.5, 0, 0)$, $u_j \sim \mathcal{N}(0, 1)$, $j = 1,\ldots, J$ and  $\epsilon \sim \mathcal{N}(0, 1)$.

\begin{figure}[!ht]
\begin{center}
\includegraphics[trim= {0.0cm 0.00cm 0.0cm 0.0cm}, clip,  width=0.99\columnwidth]{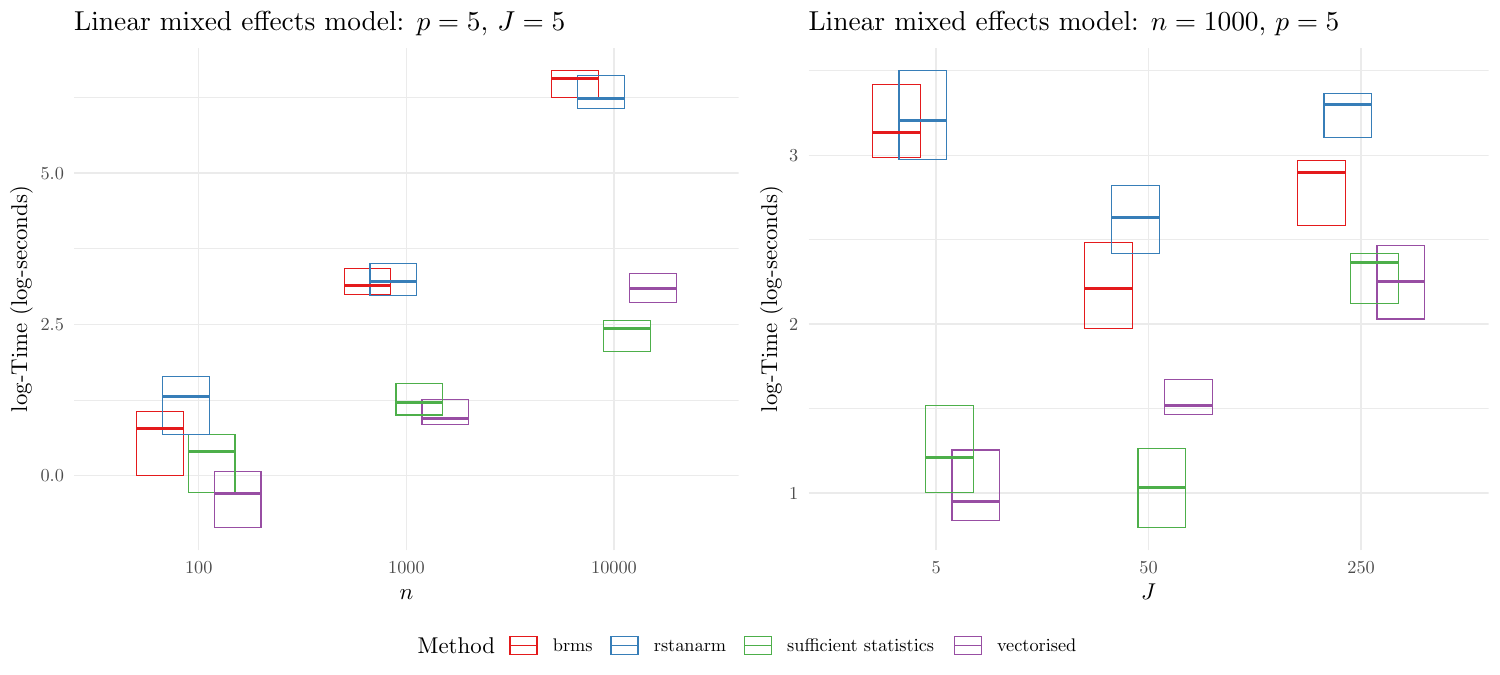}
\caption{Time taken to produce 5000 posterior samples after 1000 warm-up iterations from a Bayesian linear mixed effects model   using different \texttt{Stan} implementations. \textbf{Left}: Increasing number of observations $n$. \textbf{Right}: Increasing number of random effects $J$.
}
\label{Fig:random_effects}
\end{center}
\end{figure}

Figure \ref{Fig:random_effects} compares the time taken to produce 5000 post warm-up samples after a 1000 sample warm-up period for the different implementations repeated 25 times using the \texttt{microbench} package \citep{mersmann2024microbench}. As the number of observations $n$ increases all four methods require more time, however the time taken by the implementation taking advantage of sufficient statistics increases at the slowest rate. The sufficient statistic implementation requires more time as $n$ increase as the sufficient statistics $\bar{\bm{y}}$ and $\bar{X}$ require sums over the $n$ observation. The vectorised implementation performs surprisingly well here, but for large $n$ is slower than using sufficient statistics. As the number of groups $J$ increases the time taken by the sufficient statistics implementation as well as \texttt{brms} and \texttt{rstanarm} initially decreases before increasing. This happens as \texttt{Stan} appears to require fewer leapfrog steps and shallower NUTs trees for $J = 50$ than $J = 5$. This may be a consequence of hierarchical parameters associated to higher levels having strictly slower MCMC mixing  \citep{zanella2021multilevel}. However, the sufficient statistics implementation remains faster than the \texttt{brms} and \texttt{rstanarm} implementations throughout.






\subsection{Factor Models}{\label{Sec:FactorModels}}



Consider multivariate observations $\bm{y} = (\bm{y}_1, \ldots, \bm{y}_n)$ with $\bm{y}_i \in\mathbb{R}^p$, $i = 1,\ldots, n$. Factor models assume that there are $d < p$ latent dimensions that drive the variation in $\bm{y}$ i.e. 
\begin{equation}
    \bm{y}_i \sim \mathcal{N}_p\left(\Lambda \bm{f}_i, diag(\bm{\psi})\right), \quad \bm{f}_i \sim \mathcal{N}_p\left(\bm{0}, diag(\bm{1})\right)\nonumber
\end{equation}
where $\bm{f}_i\in\mathbb{R}^d$ are the $d$-dimensional latent factors, $\Lambda\in\mathbb{R}^{p\times d}$ is how they `load' onto $\bm{y}_i$, $\bm{\psi} \in \mathbb{R}_{+}^p$ is a vector of idiosyncratic variances and $diag(\bm{v})$ is a diagonal matrix with elements $\bm{v}$ on its diagonal. 
The full parameter vector is $\bm{\theta} = \{\Lambda, \bm{\psi}\}$.
The latent factors can be marginalised out to obtain 
\begin{equation}
    \bm{y} \sim \mathcal{N}_p\left(\bm{0}, \Lambda\Lambda^\top + diag(\psi)\right).\nonumber
\end{equation}
The multivariate Gaussian distribution is an exponential family distribution but the factor representation of the covariance prevents the use of the conjugate inverse-wishart prior. However, writing $\Omega := \left(\Lambda\Lambda^\top + diag(\bm{\psi})\right)^{-1}$, the multivariate log-likelihood can still be written as
\begin{equation}
    \ell(\bm{y}; \Lambda, \psi) = \frac{n}{2}\left(-d\log(2\pi) + |\Omega| - tr(S\Omega)\right),\nonumber
\end{equation}
where $S := Y^\top Y/n$, with $Y \in \mathbb{R}^{n\times p}$ having rows $\bm{y}_i$, is the sufficient statistic. 



We compare a vectorised extension of the \texttt{Stan} factor modelling implementation of \cite{farouni2015fitting} (Listing \ref{Lst:stan_factor_vectorised}) with an implementation of the same model that takes advantage of sufficient statistics (Listing \ref{Lst:stan_factor_suffstat}). \cite{farouni2015fitting}  impose that the upper triangular elements of $\Lambda$ are 0, and that the diagonal elements are positive to combat rotation invariance of the likelihood. They further specify hierarchical priors $\psi_j\sim \textrm{Half-Cauchy}(\mu_{\psi}, \sigma_{\psi})$, $j = 1,\ldots, p$, with $\mu_{\psi}\sim\textrm{Half-Cauchy}(0, 1)$ and $\sigma_{\psi}\sim\textrm{Half-Cauchy}(0, 1)$, $L_{jj} \sim\textrm{Half-Cauchy}(0, 3)$, $j = 1,\ldots, d$ and $L_{jk}\sim \textrm{Cauchy}(\mu_{L}, \sigma_{L})$, $j = 2,\ldots, p$, $k < d$, with $\mu_{L}\sim\textrm{Cauchy}(0, 1)$ and $\sigma_{L}\sim\textrm{Half-Cauchy}(0, 1)$.

We further consider an implementation (Listing \ref{Lst:stan_factor_SuffStat_Woodbury}) that combines sufficient statistics with a Woodbury decomposition that computes
\begin{align}
    \Omega = diag\left(\bm{\Psi}^{-1}\right) - diag\left(\bm{\Psi}^{-1}\right)\Lambda\left(diag(\bm{1}) + \Lambda^\top diag\left(\bm{\Psi}^{-1}\right)\Lambda\right)^{-1}\Lambda^\top diag\left(\bm{\Psi}^{-1}\right).\nonumber
\end{align}
While computing $\Omega$ naively requires $\mathcal{O}(p^3)$ operations, $\left(diag(\bm{1}) + \Lambda^\top diag\left(\bm{\Psi}^{-1}\right)\Lambda\right)^{-1}$ requires only $\mathcal{O}(d^3)$ \citep[see e.g.][]{ghahramani1996algorithm} and can therefore offer considerable saving for $d < p$. Figure \ref{Fig:factor_posteriors} overlays the posterior approximations generated by the different \texttt{Stan} implementations and confirms that all are sampling from the same posterior.

We generated $n \in \{100, 500, 1000\}$ observations from a factor model $\bm{y_i} \sim \mathcal{N}_p\left(\bm{0}, \Lambda\Lambda^\top + diag(\psi)\right)$ with observation dimension and number of factors $(p, d)\in \{(10, 3), (20, 5)\}$. Values for $L$ and $\bm{\psi}$ are given in Section \ref{Sec:FactorSupplement}.

\begin{figure}[!ht]
\begin{center}
\includegraphics[trim= {0.0cm 0.00cm 0.0cm 0.0cm}, clip,  width=0.99\columnwidth]{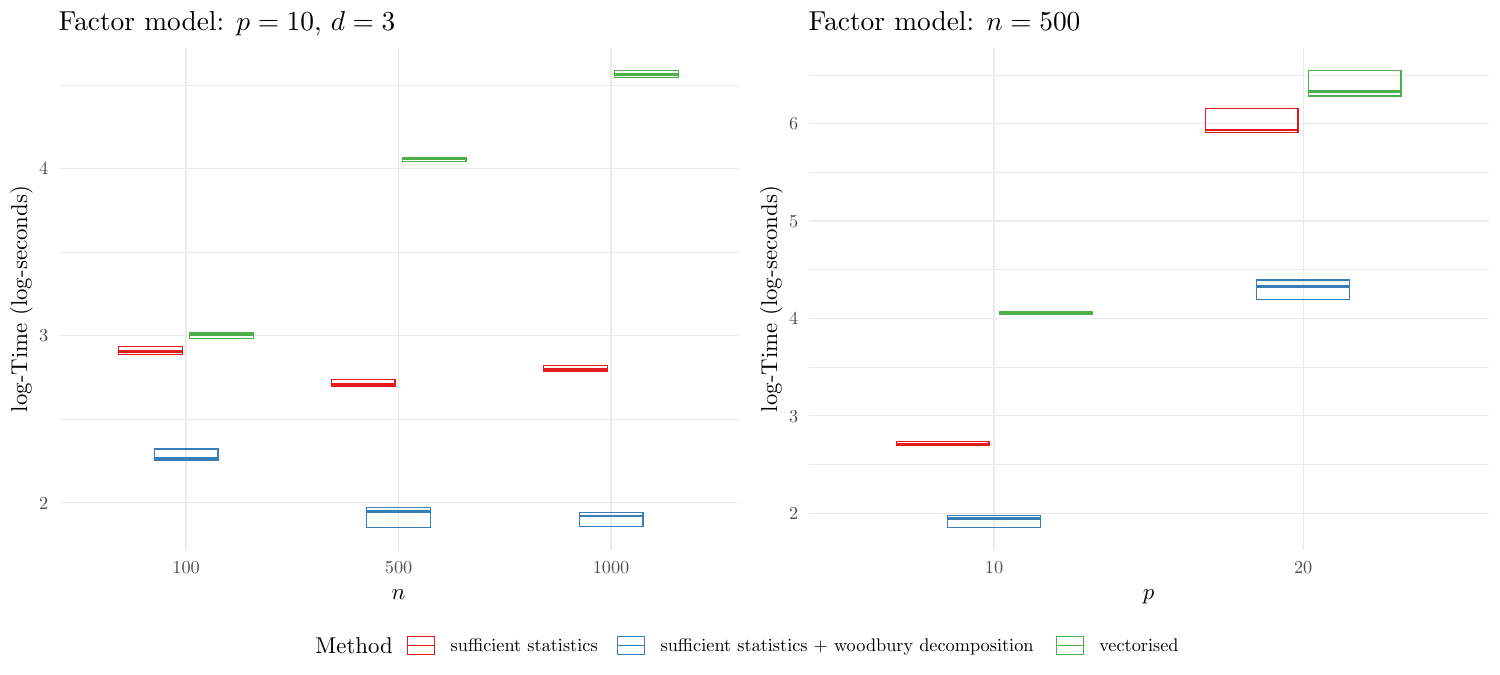}
\caption{Time taken to produce 5000 posterior samples after 1000 warm-up iterations from a Bayesian factor model using different \texttt{Stan} implementations. \textbf{Left}: Increasing number of observations $n$. \textbf{Right}: Increasing observation dimension and number of factors $p$ and $d$.}
\label{Fig:factor_analysis}
\end{center}
\end{figure}

Figure \ref{Fig:factor_analysis} compares the time taken to produce 5000 post warm-up samples after a 1000 sample warm-up period for the different implementations repeated 10 times using the \texttt{microbench} package \citep{mersmann2024microbench}. As the number of observations $n$ increases the time taken by our sufficient statistics implementations decreases slightly, while the time required by the vectorised implementation increases considerably. The decrease in time is caused by \texttt{Stan} requiring fewer leapfrog steps and shallower NUTs trees for larger $n$. Taking advantage of the Woodbury decomposition leads to a further increase in sampling speed. As $p$ and $d$ increase, the time taken for all three implementations increases, but the sufficient statistics versions remains the fastest with the saving of the Woodbury decomposition greater for larger $p$.

\subsection{Poisson Regression}{\label{Sec:Poisson}}

While massive gains can be achieved for some models, writing the likelihood entirely in terms of sufficient statistics is not always possible. For example, for observations $\bm{y} = (y_1, \ldots, y_n)\in\mathbb{N}^n$ consider a Poisson generalised linear model (GLM) where $y_i \sim Poi\left(\exp\left(\bm{x}_i^\top \bm{\beta}\right)\right)$ for predictors $\bm{x}_i\in\mathbb{R}^p$ and regression parameters $\bm{\beta}\in\mathbb{R}^p$, $i = 1,\ldots, n$. While the Poisson distribution is in the exponential family, the addition of regressors for each observation means that the log-likelihood ignoring terms that don't depend on $\beta$ only simplifies to 
\begin{equation}
    \ell(\bm{y}; \bm{\beta}) = S_{yx}^\top\bm{\beta} - \sum_{i=1}^{n}\exp(X_i \cdot \beta),\nonumber
\end{equation}
where $S_{yx} = X^\top y$ 
are partial sufficient statistics. This is only written partially in terms of sufficient statistics as the final term unavoidably contains a sum of $n$ terms that cannot be precomputed. There is no conjugate prior available for such a model and therefore MCMC is required to sample from the posterior. 

We compare a vectorised implementation (Listing \ref{Lst:stan_poisson_vectorised}), the \texttt{Stan} code produced by \texttt{brms} (Listing \ref{Lst:stan_poisson_brms}) and \texttt{rstanarm}'s implementation with an implementation that tries to take advantage of the partial sufficient statistics and vectorises the computation of  $\sum_{i=1}^{n}\exp(X_i \cdot \beta)$ (Listing \ref{Lst:stan_poisson_suffstat}). For all models, we set $\beta_j \sim \mathcal{N}(0, 2^2)$, $j=1,\ldots, p$. Figure \ref{Fig:poisson_posteriors} overlays the posterior approximations generated by the different \texttt{Stan} implementations and confirms that all are sampling from the same posterior.

We generated $n \in \{100, 1000, 10000\}$ observations from a Poisson GLM $y_{i} \sim Poi\left(\exp\left(\bm{\beta}^TX_{i}\right)\right)$ with $p \in \{10, 50, 100\}$ dimensional predictors simulated such that $X_{ij}\sim \mathcal{N}(0, 0.5)$ and regression coefficients $\beta_1 = 1.5$, $\beta_2 = 2$, $\beta_3 = 2.5$ and $\beta_j = 0$, $j = 4, \ldots, p$.

\begin{figure}[!ht]
\begin{center}
\includegraphics[trim= {0.0cm 0.00cm 0.0cm 0.0cm}, clip,  width=0.99\columnwidth]{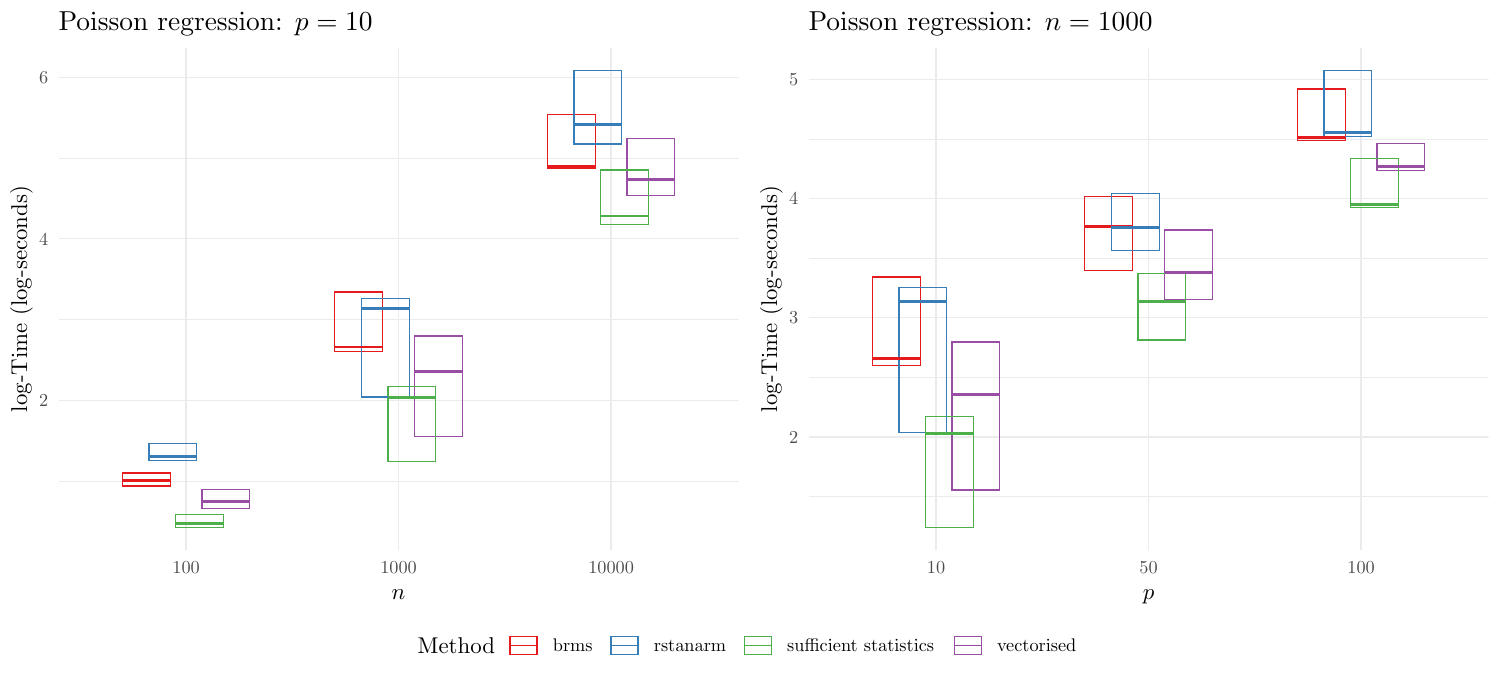}
\caption{Time taken to produce 5000 posterior samples after 1000 warm-up iterations from a Bayesian Poisson regression model using different \texttt{Stan} implementations. \textbf{Left}: Increasing number of observations $n$. \textbf{Right}: Increasing number of regression parameters $p$.}
\label{Fig:poisson_regression}
\end{center}
\end{figure}

Figure \ref{Fig:poisson_regression} compares the time taken to produce 5000 post warm-up samples after a 1000 sample warm-up period for the different implementations repeated 25 times using the \texttt{microbench} package \citep{mersmann2024microbench}. As both the number of observations, $n$, and number of regression parameters, $p$, increases, the time required for all four implementations increases at more or less than same rate. However, the time taken by the implementation taking advantage of the partial sufficient statistics is the fastest. While the computational gains of using the partial sufficient statistics as $n$ grows are less pronounced than in the previous three examples, it appear as though gains are still possible.

\section{Discussion}{\label{Sec:discussion}}

While probabilistic programming tools have made significant strides in making Bayesian inference more accessible to practitioners, there remains a tension between usability and computational efficiency. The easiest way to `write' a Bayesian model may not always correspond to the most efficient way to compute with it. 

We have demonstrated three applications where writing the likelihood in terms of its sufficient statistics provides a considerable computational improvement over current probabilistic programming implementations. While writing the full likelihood in terms of sufficient statistics is not possible for all models, it is possible for some of the most widely used models, and a further example shows that writing the likelihood partially in terms of sufficient statistics can still lead to computational gains.

We hope that the proposals made here can be incorporated into packages such as \texttt{brms} and \texttt{rstanarm} so the users can continue to benefit from their accessibility while capitalizing on the computational gains demonstrated in this paper.




\bigskip
\begin{center}
{\large\bf SUPPLEMENTARY MATERIAL}
\end{center}

Supplementary Materials contains all \texttt{Stan} code referred to in the main paper and plots showing that the different \texttt{Stan} implementations sample from the same posteriors.






\bibliographystyle{elsarticle-harv} 
\bibliography{literature.bib}

\appendix
\section{Stan models}
\label{stan_models}

We provide the \texttt{Stan} code for all models and implementations considered in the paper and demonstrate that they allow for sampling from the correct posteriors.

Listing \ref{Lst:stan_gaussian} presents example \texttt{Stan} code for sampling from a Gaussian location model with non-conjugate prior. The \texttt{model} section resembles how one would write this Bayesian model on a whiteboard.

\begin{lstlisting}[style=myListingStyle, caption=\texttt{Stan} code for a Gaussian location model with non-conjugate prior., label=Lst:stan_gaussian]
data{
  int<lower=0> n; // number of data points
  real y[n]; // observed data
}

parameters{
  real mu; // mean parameter
}

model{
  y ~ normal(mu, 1); // likelihood
  mu ~ cauchy(0, 5); // prior on mu
}
\end{lstlisting}

\subsection{Gaussian Linear Regression}

Here we provide the \texttt{Stan} code used to implement the Bayesian linear regression example in Section \ref{Sec:Regression}.

Listing \ref{Lst:stan_regression_userguide} presents a vectorised implementation as recommended in the Stan User Guide \citep{stan2024stan}.


\begin{lstlisting}[style=myListingStyle, caption=Vectorised implementation of Bayesian linear regression with non-conjugate priors in \texttt{Stan}., label=Lst:stan_regression_userguide]
data {
  int<lower=1> N;  // total number of observations
  vector[N] Y;  // response variable
  int<lower=1> K;  // number of population-level effects
  matrix[N, K] X;  // population-level design matrix
}

parameters {
  vector[K] b;  // regression coefficients
  real<lower=0> sigma;  // dispersion parameter
}

model {
  b ~ normal(0, 10);
  sigma ~ student_t(3, 0, 3.7);
  Y ~ normal(X*b, sigma);
}
\end{lstlisting}

Listing \ref{Lst:stan_regression_brms} is the \texttt{Stan} code produced by \texttt{brms} for this model.

\begin{lstlisting}[style=myListingStyle, caption= \texttt{brms} implementation of Bayesian linear regression with non conjugate priors in \texttt{Stan}., label=Lst:stan_regression_brms]
// generated with brms 2.22.0
functions {
}
data {
  int<lower=1> N;  // total number of observations
  vector[N] Y;  // response variable
  int<lower=1> K;  // number of population-level effects
  matrix[N, K] X;  // population-level design matrix
  int prior_only;  // should the likelihood be ignored?
}
transformed data {
}
parameters {
  vector[K] b;  // regression coefficients
  real<lower=0> sigma;  // dispersion parameter
}
transformed parameters {
  real lprior = 0;  // prior contributions to the log posterior
  lprior += normal_lpdf(b | 0, 10);
  lprior += student_t_lpdf(sigma | 3, 0, 3.2)
    - 1 * student_t_lccdf(0 | 3, 0, 3.2);
}
model {
  // likelihood including constants
  if (!prior_only) {
    target += normal_id_glm_lpdf(Y | X, 0, b, sigma);
  }
  // priors including constants
  target += lprior;
}
generated quantities {
}
\end{lstlisting}

Listing \ref{Lst:stan_regression_suffstat} presents our implementation of Bayesian linear regression with non-conjugate priors that leverages the sufficient statistics representation of the likelihood.

\begin{lstlisting}[style=myListingStyle, caption=Bayesian linear regression with non-conjugate priors taking advantage of sufficient statistics in \texttt{Stan}., label=Lst:stan_regression_suffstat]
data {
  int<lower=1> N;  // total number of observations
  vector[N] Y;  // response variable
  int<lower=1> K;  // number of population-level effects
  matrix[N, K] X;  // population-level design matrix
}

transformed data {
  //vector[K] means_X;  // column means of X before centering
  real Syy;
  row_vector[K] Syx;
  matrix[K,K] Sxx;

  Syy = Y'*Y;
  Syx = Y'*X;
  Sxx = crossprod(X);
}

parameters {
  vector[K] b;  // regression coefficients
  real<lower=0> sigma;  // dispersion parameter
}

transformed parameters {   
}

model {
  // Priors:
  target += normal_lpdf(b | 0, 10);
  target += student_t_lpdf(sigma | 3, 0, 3.7)
    - 1 * student_t_lccdf(0 | 3, 0, 3.7);
  // Likelihood:
  target += -N*log(sigma)-(Syy-2*Syx*b+b'*Sxx*b)/(2*sigma^2);
}
\end{lstlisting}

\begin{figure}[!ht]
\begin{center}
\includegraphics[trim= {0.0cm 0.00cm 0.0cm 0.0cm}, clip,  width=0.32\columnwidth]{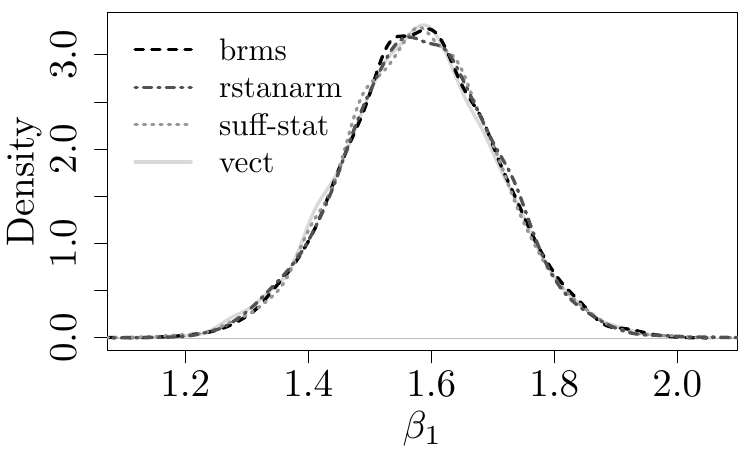}
\includegraphics[trim= {0.0cm 0.00cm 0.0cm 0.0cm}, clip,  width=0.32\columnwidth]{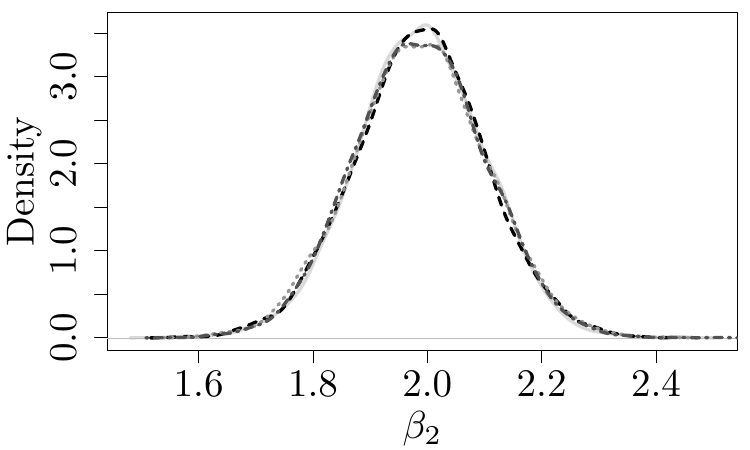}
\includegraphics[trim= {0.0cm 0.00cm 0.0cm 0.0cm}, clip,  width=0.32\columnwidth]{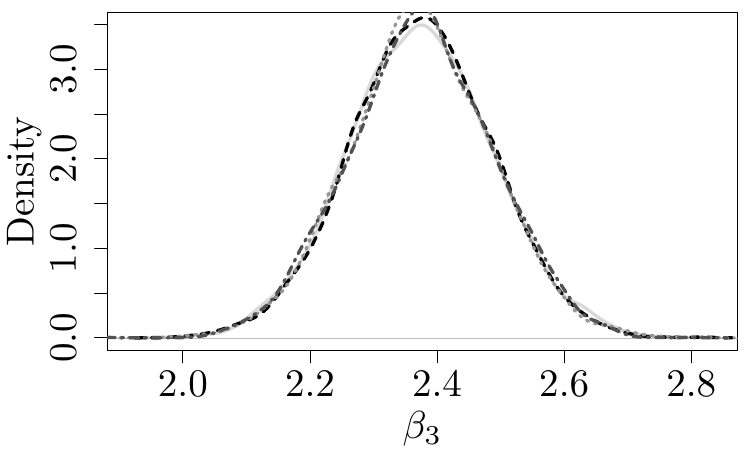}\\
\includegraphics[trim= {0.0cm 0.00cm 0.0cm 0.0cm}, clip,  width=0.32\columnwidth]{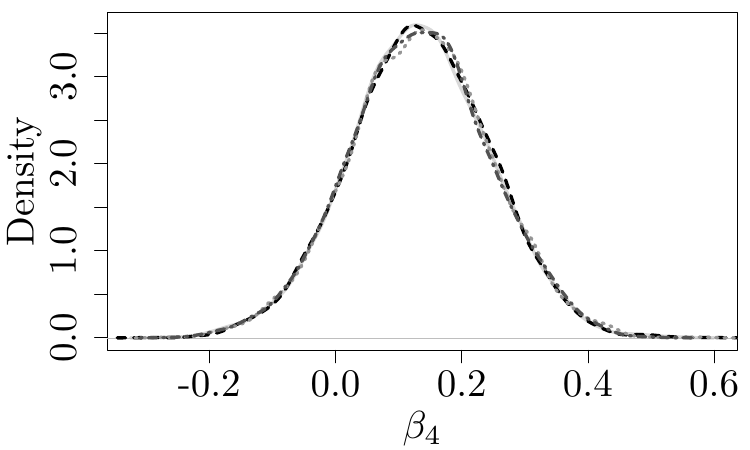}
\includegraphics[trim= {0.0cm 0.00cm 0.0cm 0.0cm}, clip,  width=0.32\columnwidth]{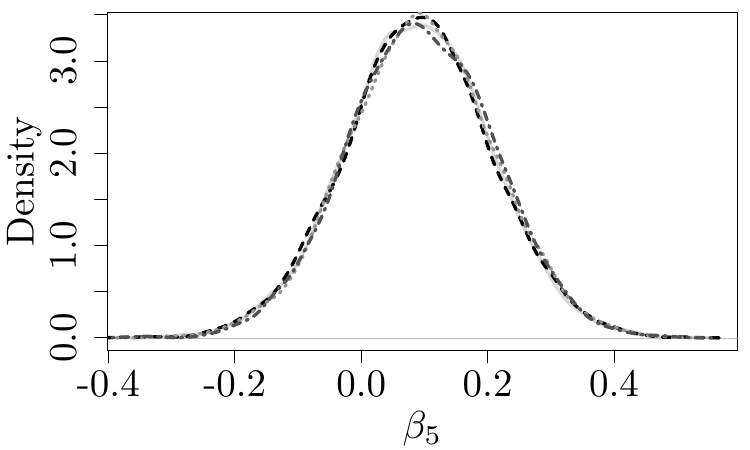}
\includegraphics[trim= {0.0cm 0.00cm 0.0cm 0.0cm}, clip,  width=0.32\columnwidth]{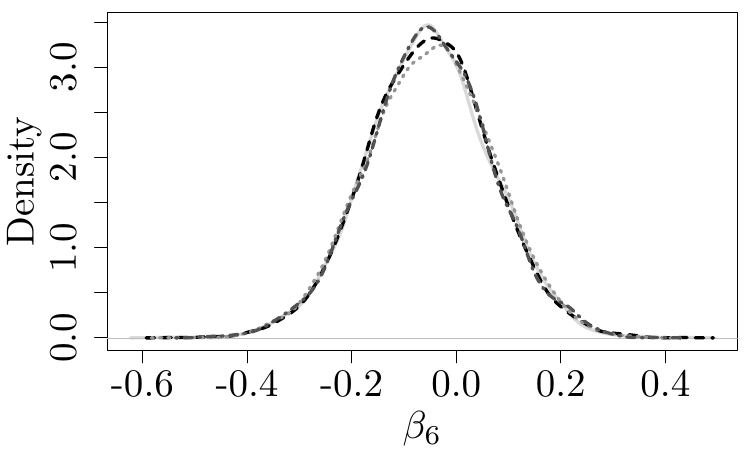}\\
\includegraphics[trim= {0.0cm 0.00cm 0.0cm 0.0cm}, clip,  width=0.32\columnwidth]{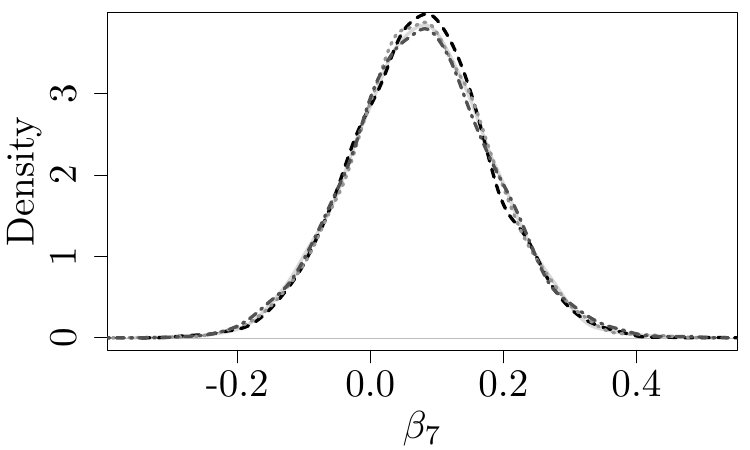}
\includegraphics[trim= {0.0cm 0.00cm 0.0cm 0.0cm}, clip,  width=0.32\columnwidth]{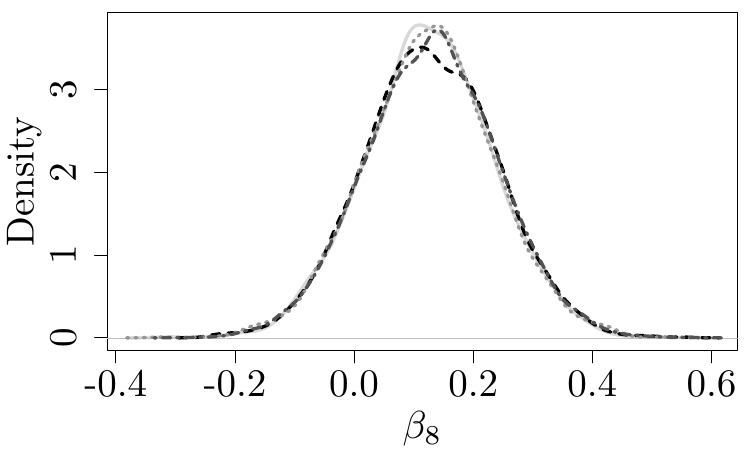}
\includegraphics[trim= {0.0cm 0.00cm 0.0cm 0.0cm}, clip,  width=0.32\columnwidth]{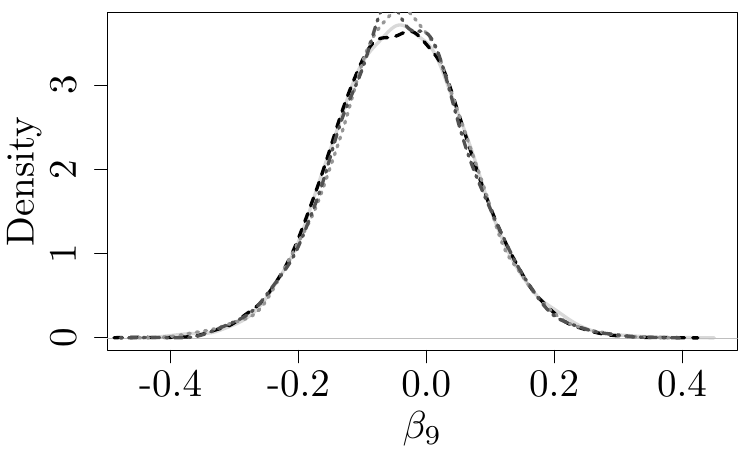}\\
\includegraphics[trim= {0.0cm 0.00cm 0.0cm 0.0cm}, clip,  width=0.32\columnwidth]{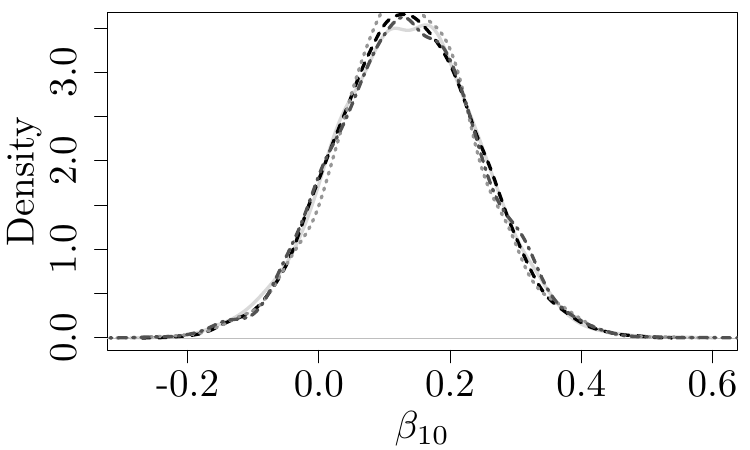}
\includegraphics[trim= {0.0cm 0.00cm 0.0cm 0.0cm}, clip,  width=0.32\columnwidth]{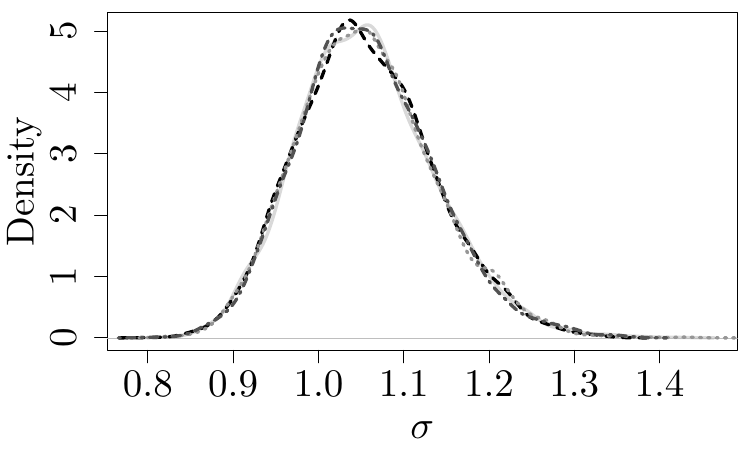}
\caption{Gaussian linear regression posteriors for $n = 100$ and $p = 10$ as estimated by different \texttt{Stan} implementations}
\label{Fig:regression_posteriors}
\end{center}
\end{figure}

Figure \ref{Fig:regression_posteriors} compares the posterior samples from the vectorised, \texttt{brms}, \texttt{rstanarm} and sufficient statistics \texttt{Stan} implementations and shows that all 4 models achieve equivalent posterior approximation.

\subsection{Mixed Effects Models}

Here we provide the \texttt{Stan} code used to implement the Bayesian linear mixed effects model example in Section \ref{Sec:RandomEffects}.

Listing \ref{Lst:stan_hierarchical_vectorised} presents a vectorised implementation.

\begin{lstlisting}[style=myListingStyle, caption=Vectorised implementation  of Bayesian linear mixed effects model in \texttt{Stan}., label=Lst:stan_hierarchical_vectorised]
data {
  int<lower=1> N;  // total number of observations
  vector[N] Y;  // response variable
  int<lower=1> K;  // number of population-level effects
  matrix[N, K] X;  // population-level design matrix
  // data for group-level effects of ID 1
  int<lower=1> N_1;  // number of grouping levels
  array[N] int<lower=1> J_1;  // grouping indicator per observation
}

parameters {
  vector[K] b;  // regression coefficients
  real<lower=0> sigma;  // dispersion parameter
  real<lower=0> sd_1;  // group-level standard deviations
  vector[N_1] z_1;  // standardized group-level effects
}

model {
  // Prior
  b ~ normal(0, 10);
  sigma ~ student_t(3, 0, 2.5);
  sd_1 ~ student_t( 3, 0, 2.5);
  
  // Random Effects
  target += normal_lpdf(z_1 | 0, sd_1);
  
  // likelihood
  target += normal_lpdf(Y | X*b + z_1[J_1], sigma);
}


\end{lstlisting}

Listing \ref{Lst:stan_hierarchical_brms} is the \texttt{Stan} code produced by \texttt{brms} for this model.

\begin{lstlisting}[style=myListingStyle, caption=\texttt{brms} implementation of Bayesian linear mixed effects model in \texttt{Stan}., label=Lst:stan_hierarchical_brms]
// generated with brms 2.22.0
functions {
}
data {
  int<lower=1> N;  // total number of observations
  vector[N] Y;  // response variable
  int<lower=1> K;  // number of population-level effects
  matrix[N, K] X;  // population-level design matrix
  // data for group-level effects of ID 1
  int<lower=1> N_1;  // number of grouping levels
  int<lower=1> M_1;  // number of coefficients per level
  array[N] int<lower=1> J_1;  // grouping indicator per observation
  // group-level predictor values
  vector[N] Z_1_1;
  int prior_only;  // should the likelihood be ignored?
}
transformed data {
}
parameters {
  vector[K] b;  // regression coefficients
  real<lower=0> sigma;  // dispersion parameter
  vector<lower=0>[M_1] sd_1;  // group-level standard deviations
  array[M_1] vector[N_1] z_1;  // standardized group-level effects
}
transformed parameters {
  vector[N_1] r_1_1;  // actual group-level effects
  real lprior = 0;  // prior contributions to the log posterior
  r_1_1 = (sd_1[1] * (z_1[1]));
  lprior += normal_lpdf(b | 0, 10);
  lprior += student_t_lpdf(sigma | 3, 0, 2.5)
    - 1 * student_t_lccdf(0 | 3, 0, 2.5);
  lprior += student_t_lpdf(sd_1 | 3, 0, 2.5)
    - 1 * student_t_lccdf(0 | 3, 0, 2.5);
}
model {
  // likelihood including constants
  if (!prior_only) {
    // initialize linear predictor term
    vector[N] mu = rep_vector(0.0, N);
    for (n in 1:N) {
      // add more terms to the linear predictor
      mu[n] += r_1_1[J_1[n]] * Z_1_1[n];
    }
    target += normal_id_glm_lpdf(Y | X, mu, b, sigma);
  }
  // priors including constants
  target += lprior;
  target += std_normal_lpdf(z_1[1]);
}
generated quantities {
}

\end{lstlisting}

Listing \ref{Lst:stan_hierarchical_SuffStat} presents our implementation of Bayesian linear mixed effects model that leverages the sufficient statistics representation of the likelihood.

\begin{lstlisting}[style=myListingStyle, caption=Bayesian linear mixed effects model taking advantage of sufficient statistics in \texttt{Stan}., label=Lst:stan_hierarchical_SuffStat]
functions {
}
data {
  int<lower=1> N;  // total number of observations
  vector[N] Y;  // response variable
  int<lower=1> K;  // number of population-level effects
  matrix[N, K] X;  // population-level design matrix
  // data for group-level effects of ID 1
  int<lower=1> N_1;  // number of grouping levels
  array[N] int<lower=1> J_1;  // grouping indicator per observation
}
transformed data {
  real Syy;              // Y' * Y
  row_vector[K] Syx;     // Y' * X
  matrix[K, K] Sxx;      // X' * X
  vector[N_1] u_count;   // Number of observations in each group
  vector[N_1] u_sumY;    // Sum of Y for each group
  matrix[N_1, K] u_sumX; // Sum of X for each group

  
  Syy = dot_self(Y);     // Equivalent to Y' * Y
  Syx = Y' * X;          // Equivalent to Y' * X
  Sxx = crossprod(X);    // Equivalent to X' * X
  
  u_count = rep_vector(0.0, N_1);
  u_sumY = rep_vector(0.0, N_1);
  u_sumX = rep_matrix(0.0, N_1, K);
  
  for (n in 1:N) {
    u_count[J_1[n]] += 1;
    u_sumY[J_1[n]] += Y[n];
    u_sumX[J_1[n], ] += X[n, ];
  }


}
parameters {
  vector[K] b;  // regression coefficients
  real<lower=0> sigma;  // dispersion parameter
  real<lower=0> sd_1;  // group-level standard deviations
  vector[N_1] z_1;  // standardized group-level effects
}
transformed parameters {
  vector[N_1] r_1_1;  // actual group-level effects
  r_1_1 = (sd_1 * (z_1));
  
}

model {
  // likelihood including constants

  // Adjust sufficient statistics for the group-level effects
  real Syy_adjusted = Syy - 2 * r_1_1' * u_sumY + 
                      (r_1_1^2)' * u_count;
  row_vector[K] Syx_adjusted = Syx - r_1_1' * u_sumX;

  // Likelihood using sufficient statistics
  target += -N*log(sigma) - (Syy_adjusted - 2 * Syx_adjusted * b + 
             b' * Sxx * b) / (2 * sigma^2);

  // priors including constants
  target += normal_lpdf(b | 0, 10);
  target += student_t_lpdf(sigma | 3, 0, 2.5)
    - 1 * student_t_lccdf(0 | 3, 0, 2.5);
  target += student_t_lpdf(sd_1 | 3, 0, 2.5)
    - 1 * student_t_lccdf(0 | 3, 0, 2.5);
  target += std_normal_lpdf(z_1);
}
generated quantities {
}
\end{lstlisting}

\begin{figure}[!ht]
\begin{center}
\includegraphics[trim= {0.0cm 0.00cm 0.0cm 0.0cm}, clip,  width=0.32\columnwidth]{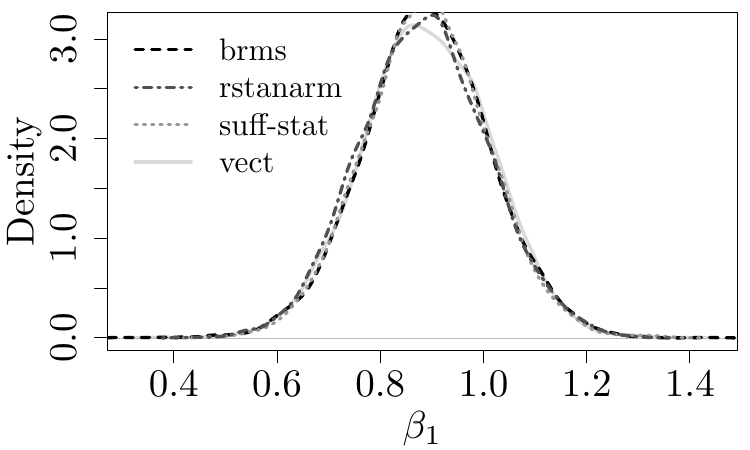}
\includegraphics[trim= {0.0cm 0.00cm 0.0cm 0.0cm}, clip,  width=0.32\columnwidth]{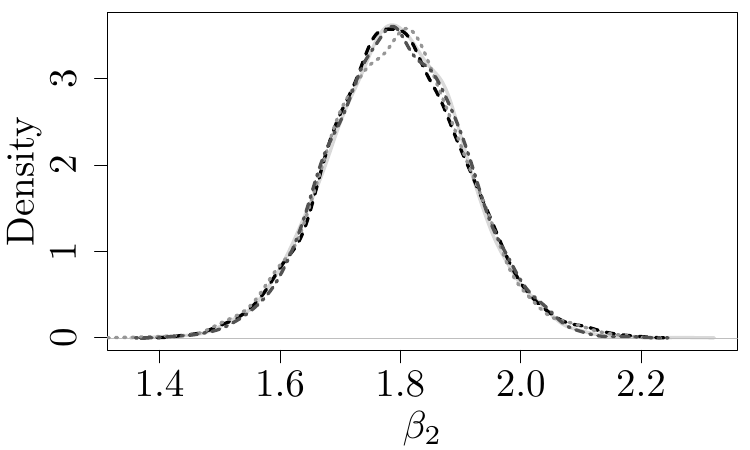}
\includegraphics[trim= {0.0cm 0.00cm 0.0cm 0.0cm}, clip,  width=0.32\columnwidth]{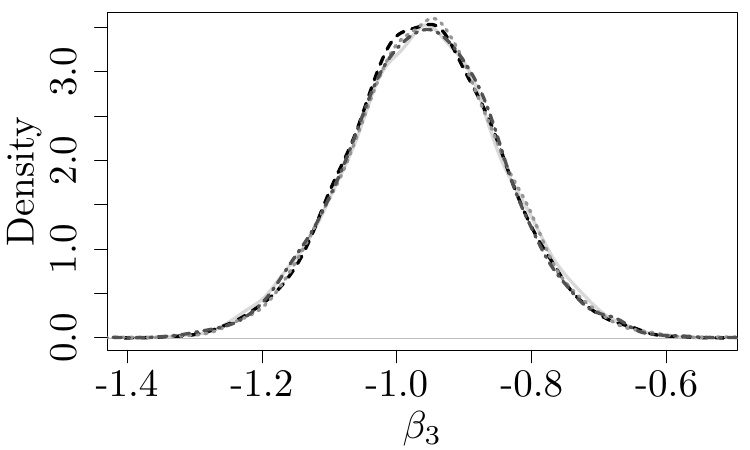}\\
\includegraphics[trim= {0.0cm 0.00cm 0.0cm 0.0cm}, clip,  width=0.32\columnwidth]{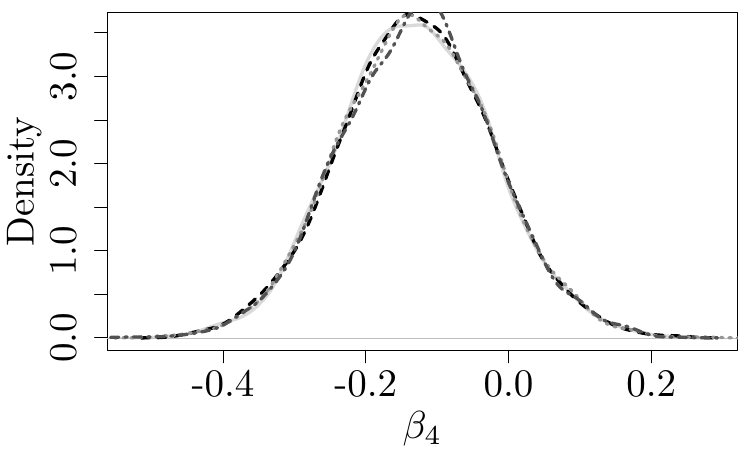}
\includegraphics[trim= {0.0cm 0.00cm 0.0cm 0.0cm}, clip,  width=0.32\columnwidth]{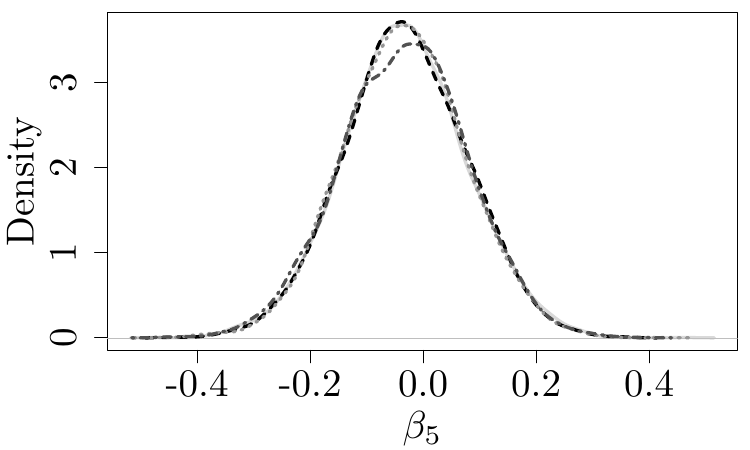}
\includegraphics[trim= {0.0cm 0.00cm 0.0cm 0.0cm}, clip,  width=0.32\columnwidth]{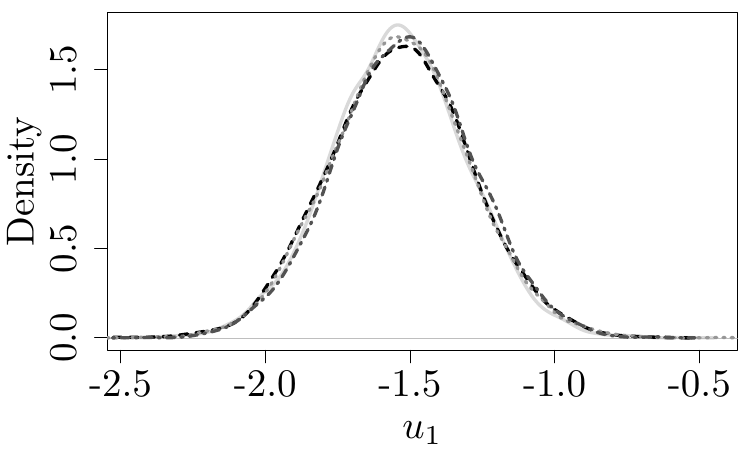}\\
\includegraphics[trim= {0.0cm 0.00cm 0.0cm 0.0cm}, clip,  width=0.32\columnwidth]{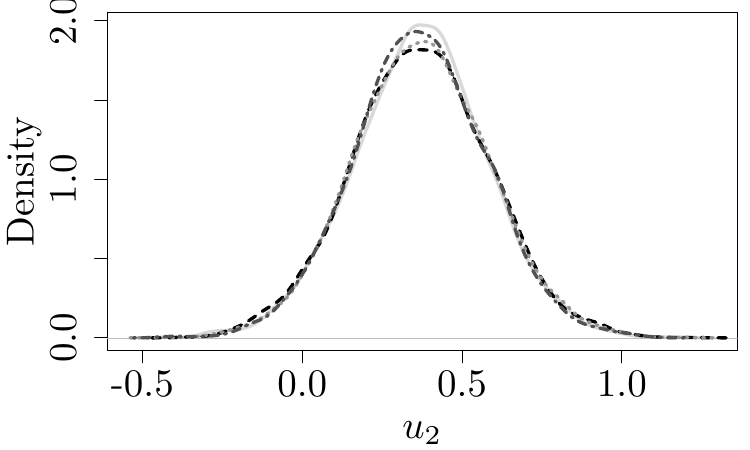}
\includegraphics[trim= {0.0cm 0.00cm 0.0cm 0.0cm}, clip,  width=0.32\columnwidth]{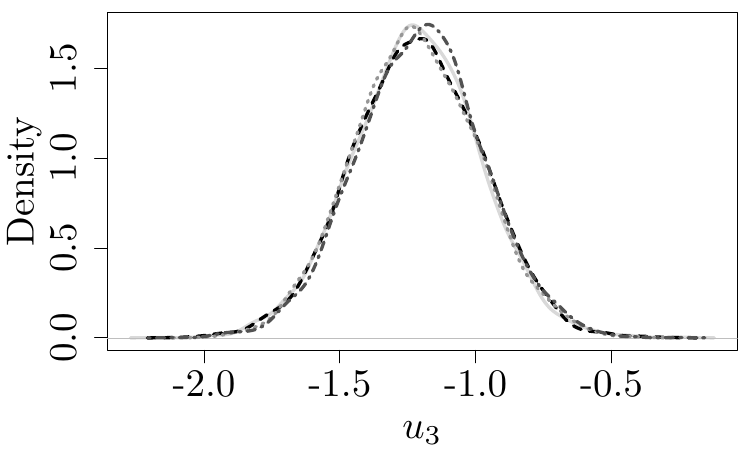}
\includegraphics[trim= {0.0cm 0.00cm 0.0cm 0.0cm}, clip,  width=0.32\columnwidth]{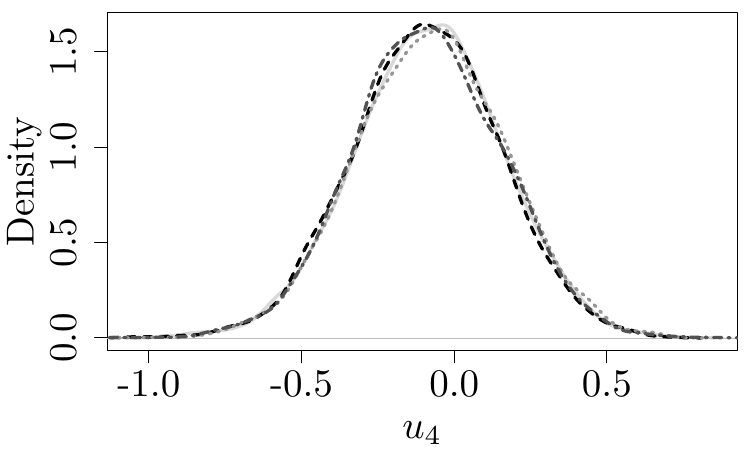}\\
\includegraphics[trim= {0.0cm 0.00cm 0.0cm 0.0cm}, clip,  width=0.32\columnwidth]{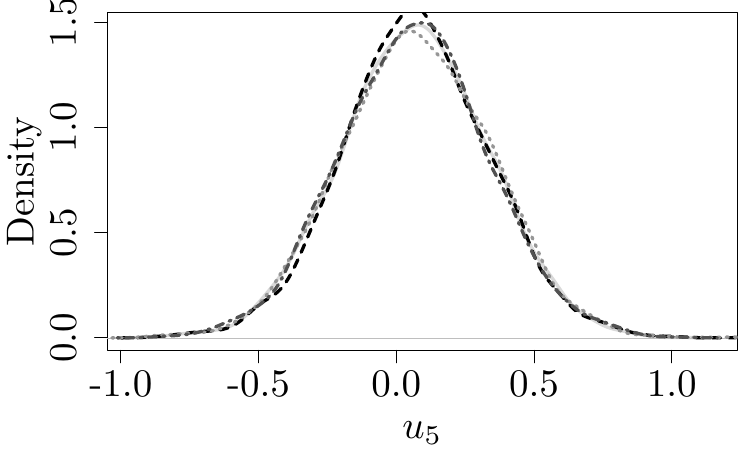}
\includegraphics[trim= {0.0cm 0.00cm 0.0cm 0.0cm}, clip,  width=0.32\columnwidth]{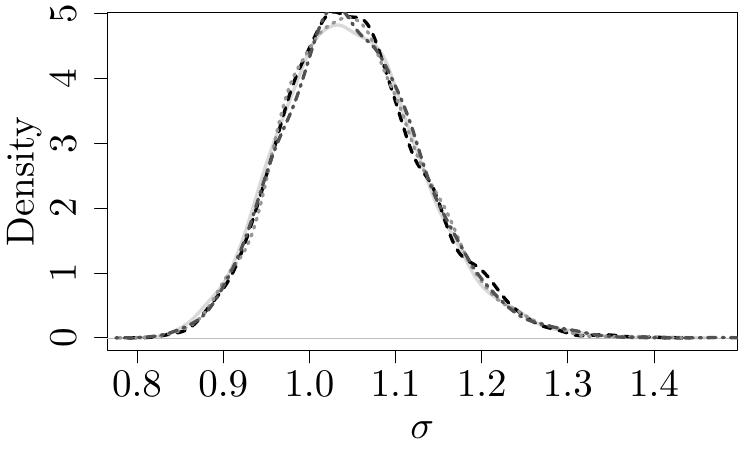}
\includegraphics[trim= {0.0cm 0.00cm 0.0cm 0.0cm}, clip,  width=0.32\columnwidth]{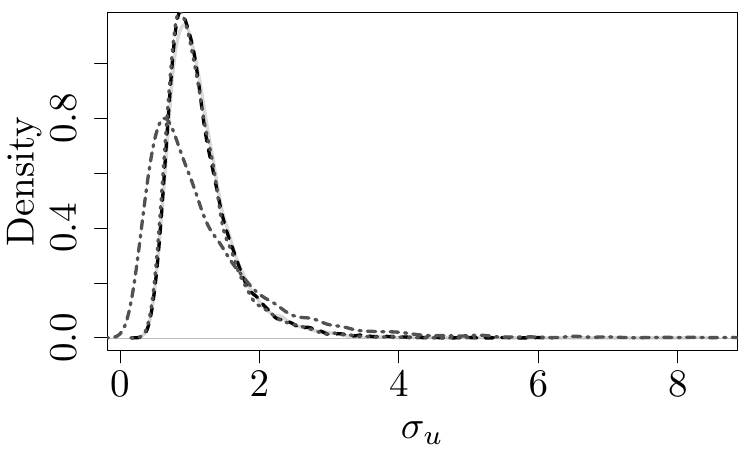}
\caption{Linear mixed effects model posteriors for $n = 100$, $p = 4$, $J = 5$ as estimated by different \texttt{Stan} implementations.}
\label{Fig:mixed_posteriors}
\end{center}
\end{figure}

Figure \ref{Fig:mixed_posteriors} compares the posterior samples from the vectorised, \texttt{brms}, \texttt{rstanarm} and sufficient statistics \texttt{Stan} implementations and shows that all 4 models achieve equivalent posterior approximation with the exception of the posterior for $\sigma_u$ produced by \texttt{rstanarm}.

\subsection{Factor Models}{\label{Sec:FactorSupplement}}

Here we provide the simulation scenario and the \texttt{Stan} code used to implement the Bayesian factor model example in Section \ref{Sec:FactorModels}.

The simulation scenarios where $p = 10$ and $d = 3$ used

{
\footnotesize
\begin{align}
L = \begin{pmatrix}
 0.99 & 0.00&0.00 \\
 0.00 & 0.90 &0.00 \\
 0.25 & 0.25 &0.85 \\
 0.00 & 0.40 &0.80 \\
 0.80 & 0.00 &0.00 \\
 0.00 & 0.50& 0.75 \\
 0.50 & 0.00 &0.75 \\
 0.00 & 0.00 &0.00 \\
 0.00 &-0.30 &0.80 \\
 0.00 &-0.30 &0.80 \\
\end{pmatrix}\quad \bm{\psi} = \begin{pmatrix}
    0.2079\\ 0.19\\ 0.1525\\ 0.20\\ 0.36\\ 0.1875\\ 0.1875\\ 1.00\\ 0.27\\ 0.27\\
\end{pmatrix}.\nonumber
\end{align}
}
The simulation scenarios where $p = 20$ and $d = 5$ used
{
\footnotesize
\begin{align}
L = \begin{pmatrix}
 -1.33 & 0.00 & 0.00 & 0.00 & 0.00 \\
 -1.22 &-0.43 & 0.00 & 0.00 & 0.00 \\
  0.66 &-1.21 & 0.34 & 0.00 & 0.00 \\
 -0.15 & 0.02 & 0.04 &-0.27 & 0.00 \\
 -0.89 & 0.10 &-1.50 & 0.29 &-0.24 \\
 -0.09 &-0.18 & 0.14 & 0.38 & 0.33 \\
  0.61 & 0.38 &-0.18 & 0.23 & 0.70 \\
  0.95 &-0.36 & 0.09 &-0.44 &-0.56 \\
 -0.22 &-0.68 & 0.29 &-0.55 &-0.43 \\
 -0.13 & 0.22 & 0.70 & 0.76 &-0.57 \\
 -0.88 &-0.41 &-0.36 & 0.13 &-0.73 \\
  0.23 & 0.72 & 0.65 & 0.04 & 0.04 \\
 -0.32 &-0.22 & 0.17 &-0.06 & 0.33 \\
  0.23 & 0.33 & 0.52 &-0.60 & 0.60 \\
  0.35 & 0.16 & 0.46 & 0.31 & 0.52 \\
  0.52 &-0.39 & 0.36 &-0.11 &-0.50 \\
 -0.30 & 0.79 &-0.52 &-0.09 & 0.92 \\
  0.25 & 0.32 &-0.05 & 0.47 &-0.33 \\
 -0.86 & 0.04 & 0.31 & 0.41 & 0.05 \\
 -0.39 & 0.14 &-0.48 & 0.70 &-0.21 \\
\end{pmatrix}\quad \bm{\psi} = \begin{pmatrix}
    0.198\\ 0.661\\ 0.283\\ 0.038\\ 0.473\\ 1.464\\ 0.314\\ 0.410\\ 1.192\\ 0.715\\ 1.345\\ 2.409\\ 0.096\\ 0.057\\ 1.254\\ 0.310\\ 0.475\\ 0.622\\ 1.246\\ 0.370\\
\end{pmatrix}.\nonumber
\end{align}
}

Listing \ref{Lst:stan_factor_vectorised} presents a vectorised extension of the implementation of \cite{farouni2015fitting}.

\begin{lstlisting}[style=myListingStyle, caption=Vectorised implementation of Bayesian factor model in \texttt{Stan}., label=Lst:stan_factor_vectorised]

data {
  int<lower=1> N;                // number of observations
  int<lower=1> P;                // dimension of observations
  vector[P] Y[N];                 // data matrix of order [N,P]
  int<lower=1> D;              // number of latent dimensions 
}

transformed data {
  int<lower=1> M;
  vector[P] mu;
  M  = D*(P-D)+ D*(D-1)/2;  // number of non-zero loadings
  mu = rep_vector(0.0,P);
}

parameters {    
  vector[M] L_t;   // lower diagonal elements of L
  vector<lower=0>[D] L_d;   // lower diagonal elements of L
  vector<lower=0>[P] psi;         // vector of variances
  real<lower=0>   mu_psi;
  real<lower=0>  sigma_psi;
  real   mu_lt;
  real<lower=0>  sigma_lt;
}
transformed parameters{
  cholesky_factor_cov[P,D] L; //lower triangular factor loadings
  cov_matrix[P] Q;   //Covariance mat
{
  int idx2 = 0;
  for(i in 1:P){
    for(j in (i+1):D){
      L[i,j] = 0; //constrain the upper triangular elements to zero 
    }
  }
  for (j in 1:D) {
      L[j,j] = L_d[j];
    for (i in (j+1):P) {
      idx2 += 1;
      L[i,j] = L_t[idx2];
    } 
  }
} 
  Q = L*L' + diag_matrix(psi); 

}
model {
   // the hyperpriors 
   target +=  cauchy_lpdf(mu_psi | 0, 1);
   target +=  cauchy_lpdf(sigma_psi | 0,1);
   target +=  cauchy_lpdf(mu_lt | 0, 1);
   target +=  cauchy_lpdf(sigma_lt | 0,1);
   // the priors 
   target +=  cauchy_lpdf(L_d | 0,3);
   target +=  cauchy_lpdf(L_t | mu_lt,sigma_lt);
   target +=  cauchy_lpdf(psi | mu_psi,sigma_psi);
   //The likelihood
   
   target += multi_normal_lpdf(Y | mu, Q); 
}

\end{lstlisting}

Listing \ref{Lst:stan_factor_suffstat} presents our implementation of Bayesian factor model that leverages the sufficient statistics representation of the likelihood.

\begin{lstlisting}[style=myListingStyle, caption=Bayesian factor model taking advantage of sufficient statistics in \texttt{Stan}., label=Lst:stan_factor_suffstat]
data {
  int<lower=1> N;                // number of 
  int<lower=1> P;                // number of 
  matrix[N,P] Y;                 // data matrix of order [N,P]
  int<lower=1> D;              // number of latent dimensions 
}

transformed data {
  int<lower=1> M;
  vector[P] mu;
  // Sufficient Statistics
  matrix[P, P] S_bar; 
  M  = D*(P-D)+ D*(D-1)/2;  // number of non-zero loadings
  mu = rep_vector(0.0,P);
  S_bar = Y'*Y/N;
}

parameters {    
  vector[M] L_t;   // lower diagonal elements of L
  vector<lower=0>[D] L_d;   // lower diagonal elements of L
  vector<lower=0>[P] psi;         // vector of variances
  real<lower=0>   mu_psi;
  real<lower=0>  sigma_psi;
  real   mu_lt;
  real<lower=0>  sigma_lt;
}
transformed parameters{
  cholesky_factor_cov[P,D] L; //lower triangular factor loadings
  cov_matrix[P] Omega;   //precision mat
{  
  int idx2 = 0;
  for(i in 1:P){
    for(j in (i+1):D){
      L[i,j] = 0; //constrain the upper triangular elements to zero 
    }
  }
  for (j in 1:D) {
      L[j,j] = L_d[j];
    for (i in (j+1):P) {
      idx2 += 1;
      L[i,j] = L_t[idx2];
    } 
  }

}
  Omega = inverse_spd(L*L'+diag_matrix(psi)); 

}
model {
   // the hyperpriors 
   target += cauchy_lpdf(mu_psi | 0, 1);
   target += cauchy_lpdf(sigma_psi |0, 1);
   target += cauchy_lpdf(mu_lt | 0, 1);
   target += cauchy_lpdf(sigma_lt | 0, 1);
   // the priors 
   target += cauchy_lpdf(L_d | 0,3);
   target += cauchy_lpdf(L_t | mu_lt, sigma_lt);
   target += cauchy_lpdf(psi | mu_psi, sigma_psi);
   //The likelihood
   // non-zero mean
   //target += N*(-0.5*P*log(2*pi()) + 
               0.5*log_determinant(Omega)  - 
               0.5*(x_bar - mu)'*Omega*(x_bar - mu) - 
               0.5*trace(S_bar*Omega));
   // zero-mean
   target += 0.5*N*(-P*log(2*pi()) + log_determinant(Omega) - 
             trace(S_bar*Omega));
}
\end{lstlisting}

Listing \ref{Lst:stan_factor_SuffStat_Woodbury} presents our implementation of Bayesian factor model that leverages the sufficient statistics representation of the likelihood and the Woodbury decomposition to invert the factor covariance matrix $\Omega$.

\begin{lstlisting}[style=myListingStyle, caption=Bayesian factor model taking advantage of sufficient statistics and the Woodbury decomposition in \texttt{Stan}., label=Lst:stan_factor_SuffStat_Woodbury]
functions {
   // Woodbury Identity
   matrix woodbury_inverse_broadcast(vector Psi_diag, matrix U){
      // Psi_diag is a p x 1 vector of the diagonal elements of Psi
      // U is a p x k matrix 
      // V is a k times p matrix
      // V = U'
      
      int dimensions[2] = dims(U);
      int p = dimensions[1];
      int k = dimensions[2];
      matrix[k, p] V = U';
      matrix[p, k] Psi_inv_broadcast = rep_matrix((1 ./ Psi_diag), k);
      matrix[k, k] B_inv = inverse(diag_matrix(rep_vector(1, k)) + 
                          V*(Psi_inv_broadcast .* U));
      
      return (diag_matrix(Psi_inv_broadcast[,1]) - 
             (Psi_inv_broadcast .* U) * (Psi_inv_broadcast .* 
             (B_inv * V)')');
   }
   
}

data {
  int<lower=1> N;                // number of 
  int<lower=1> P;                // number of 
  matrix[N,P] Y;                 // data matrix of order [N,P]
  int<lower=1> D;              // number of latent dimensions 
}

transformed data {
  int<lower=1> M;
  vector[P] mu;
  // Sufficient Statistics
  matrix[P, P] S_bar; 
  M  = D*(P-D)+ D*(D-1)/2;  // number of non-zero loadings
  mu = rep_vector(0.0,P);
  S_bar = Y'*Y/N;
}

parameters {    
  vector[M] L_t;   // lower diagonal elements of L
  vector<lower=0>[D] L_d;   // lower diagonal elements of L
  vector<lower=0>[P] psi;         // vector of variances
  real<lower=0>   mu_psi;
  real<lower=0>  sigma_psi;
  real   mu_lt;
  real<lower=0>  sigma_lt;
}
transformed parameters{
  cholesky_factor_cov[P,D] L; //lower triangular factor loadings 
  cov_matrix[P] Omega;   //precision mat
{
  int idx2 = 0;
  for(i in 1:P){
    for(j in (i+1):D){
      L[i,j] = 0; //constrain the upper triangular elements to zero 
    }
  }
  for (j in 1:D) {
      L[j,j] = L_d[j];
    for (i in (j+1):P) {
      idx2 += 1;
      L[i,j] = L_t[idx2];
    } 
  } 
}
  Omega = woodbury_inverse_broadcast(psi, L);

}
model {
   // the hyperpriors 
   target += cauchy_lpdf(mu_psi | 0, 1);
   target += cauchy_lpdf(sigma_psi |0, 1);
   target += cauchy_lpdf(mu_lt | 0, 1);
   target += cauchy_lpdf(sigma_lt | 0, 1);
   // the priors 
   target += cauchy_lpdf(L_d | 0,3);
   target += cauchy_lpdf(L_t | mu_lt, sigma_lt);
   target += cauchy_lpdf(psi | mu_psi, sigma_psi);
   //The likelihood
   target += 0.5*N*(-P*log(2*pi()) + 
             log_determinant(Omega) - 
             trace(S_bar*Omega));
}
\end{lstlisting}

\begin{figure}[!ht]
\begin{center}
\includegraphics[trim= {0.0cm 0.00cm 0.0cm 0.0cm}, clip,  width=0.24\columnwidth]{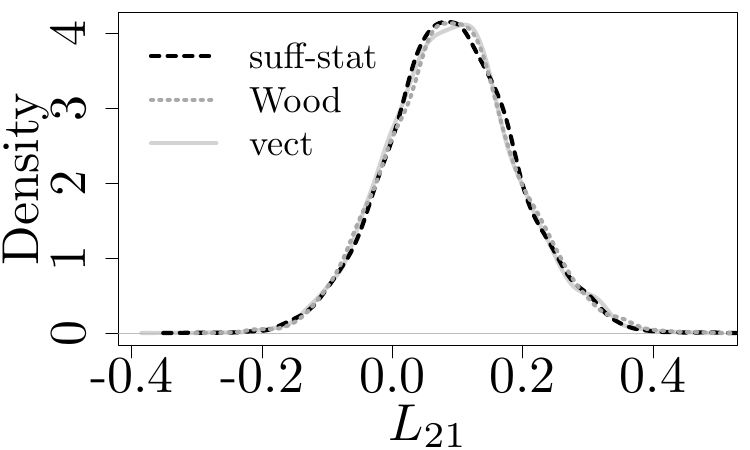}
\includegraphics[trim= {0.0cm 0.00cm 0.0cm 0.0cm}, clip,  width=0.24\columnwidth]{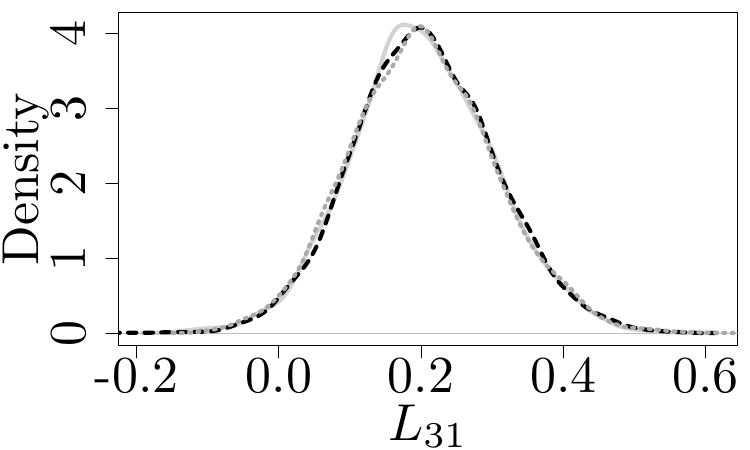}
\includegraphics[trim= {0.0cm 0.00cm 0.0cm 0.0cm}, clip,  width=0.24\columnwidth]{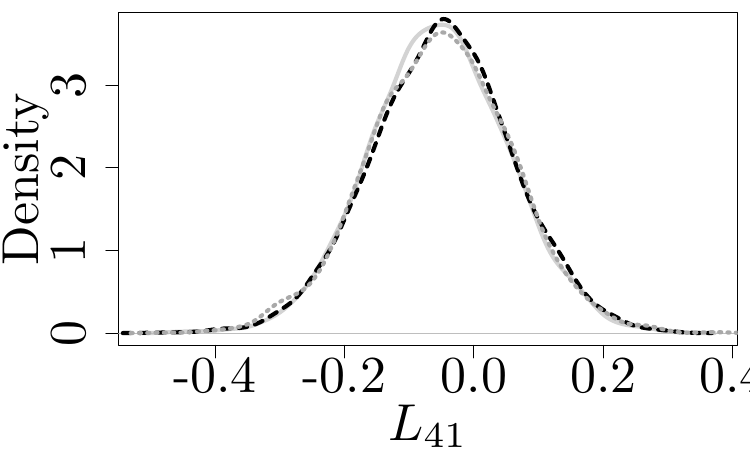}
\includegraphics[trim= {0.0cm 0.00cm 0.0cm 0.0cm}, clip,  width=0.24\columnwidth]{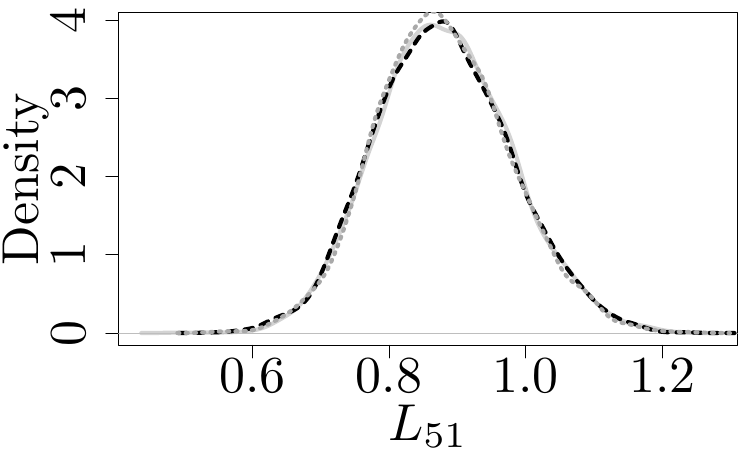}\\
\includegraphics[trim= {0.0cm 0.00cm 0.0cm 0.0cm}, clip,  width=0.24\columnwidth]{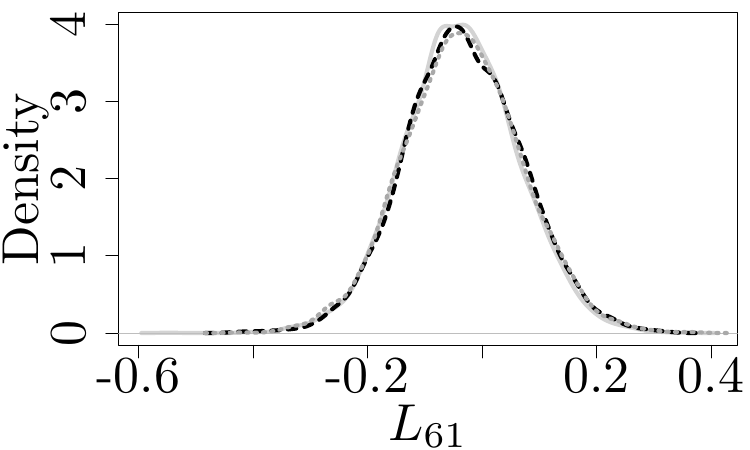}
\includegraphics[trim= {0.0cm 0.00cm 0.0cm 0.0cm}, clip,  width=0.24\columnwidth]{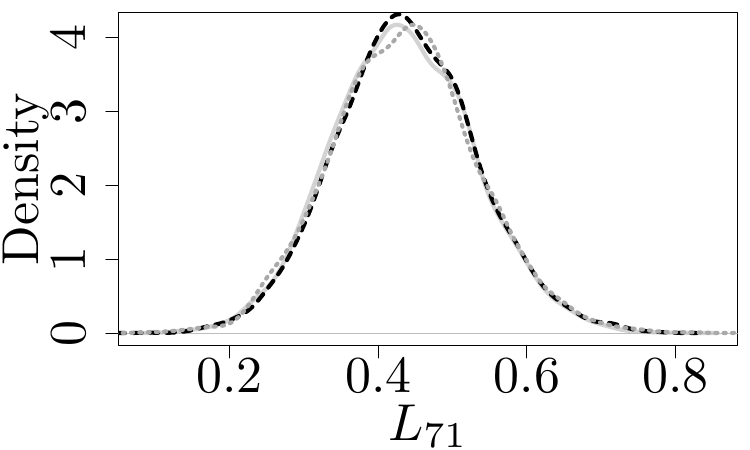}
\includegraphics[trim= {0.0cm 0.00cm 0.0cm 0.0cm}, clip,  width=0.24\columnwidth]{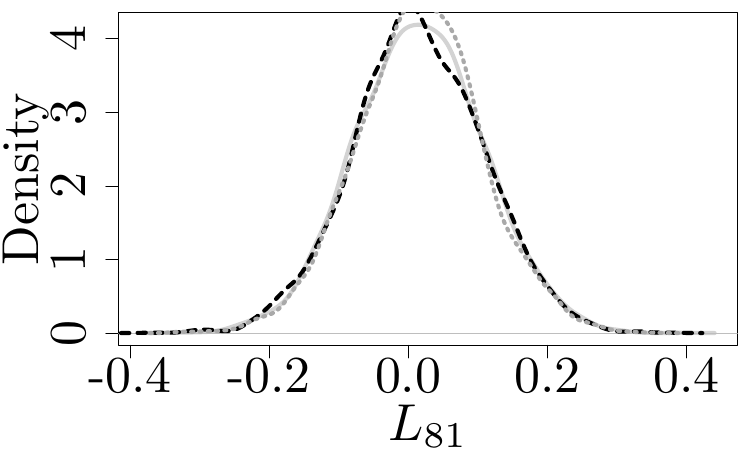}
\includegraphics[trim= {0.0cm 0.00cm 0.0cm 0.0cm}, clip,  width=0.24\columnwidth]{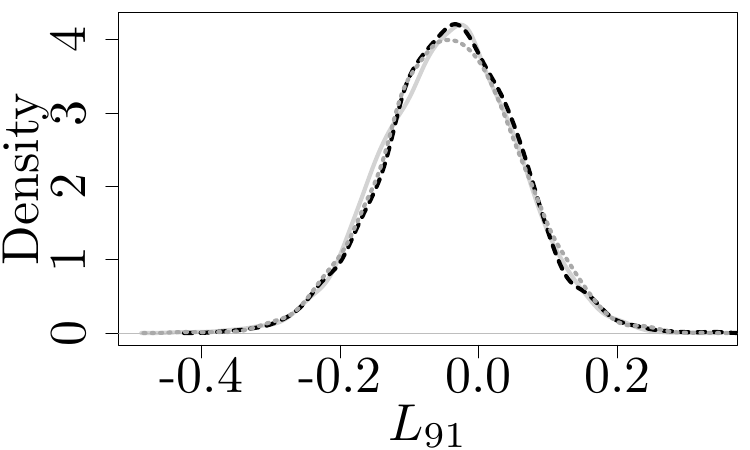}\\
\includegraphics[trim= {0.0cm 0.00cm 0.0cm 0.0cm}, clip,  width=0.24\columnwidth]{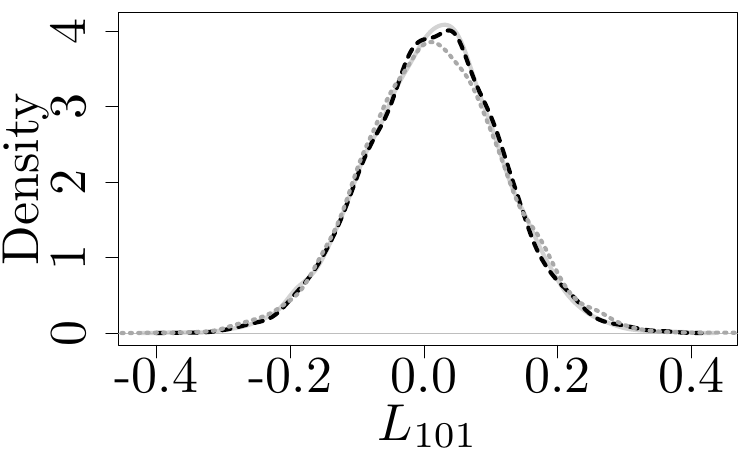}
\includegraphics[trim= {0.0cm 0.00cm 0.0cm 0.0cm}, clip,  width=0.24\columnwidth]{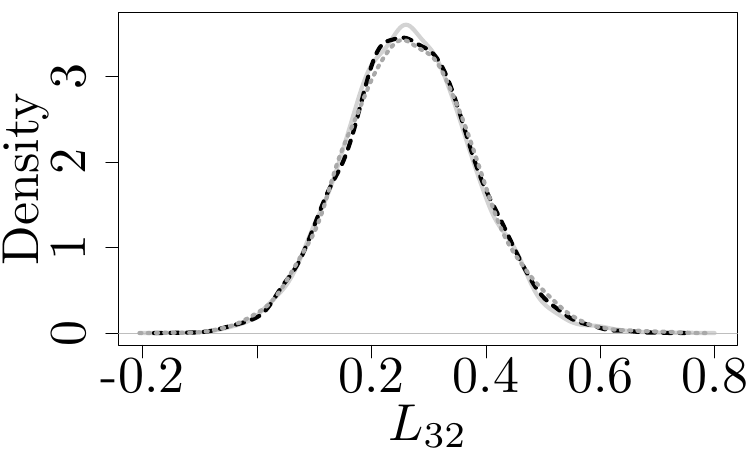}
\includegraphics[trim= {0.0cm 0.00cm 0.0cm 0.0cm}, clip,  width=0.24\columnwidth]{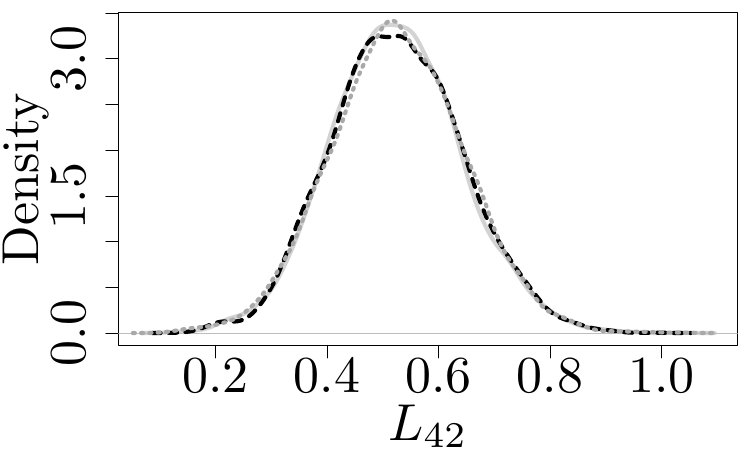}
\includegraphics[trim= {0.0cm 0.00cm 0.0cm 0.0cm}, clip,  width=0.24\columnwidth]{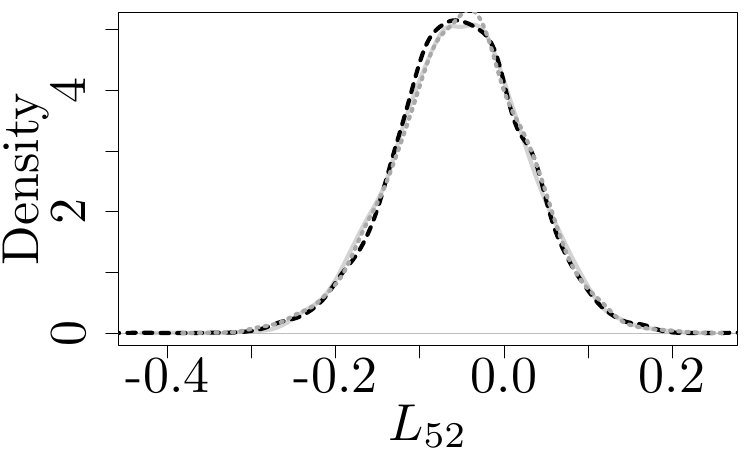}\\
\includegraphics[trim= {0.0cm 0.00cm 0.0cm 0.0cm}, clip,  width=0.24\columnwidth]{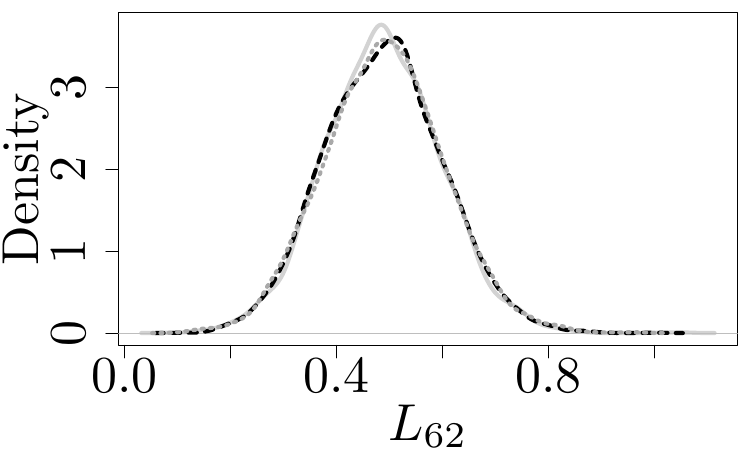}
\includegraphics[trim= {0.0cm 0.00cm 0.0cm 0.0cm}, clip,  width=0.24\columnwidth]{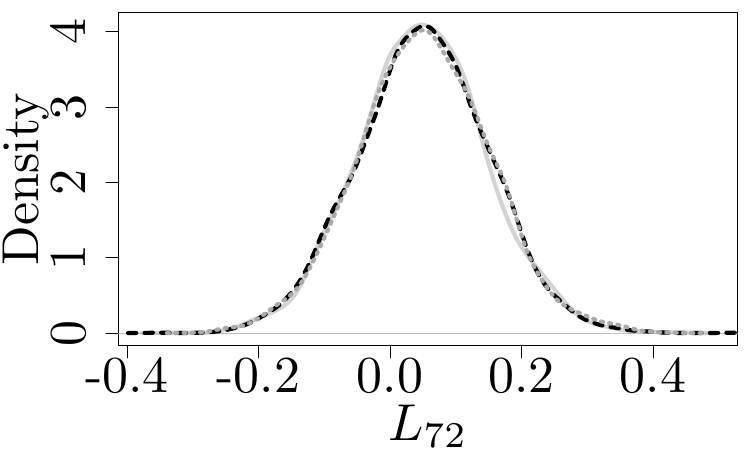}
\includegraphics[trim= {0.0cm 0.00cm 0.0cm 0.0cm}, clip,  width=0.24\columnwidth]{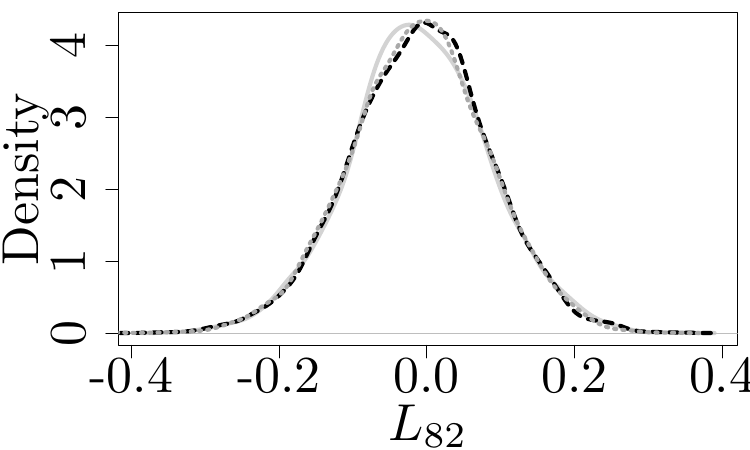}
\includegraphics[trim= {0.0cm 0.00cm 0.0cm 0.0cm}, clip,  width=0.24\columnwidth]{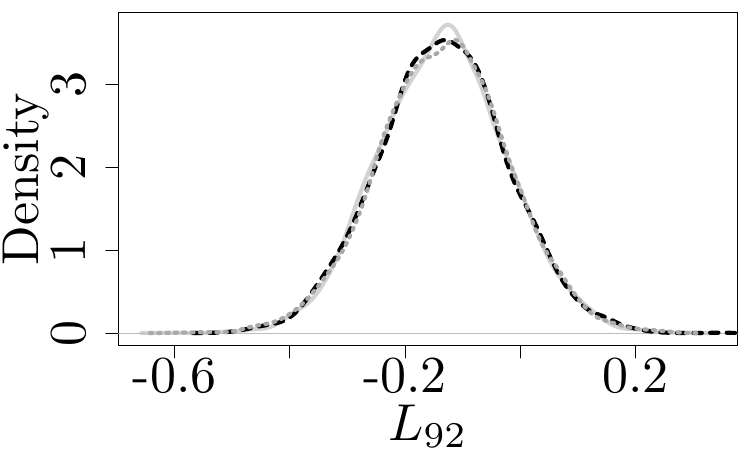}\\
\includegraphics[trim= {0.0cm 0.00cm 0.0cm 0.0cm}, clip,  width=0.24\columnwidth]{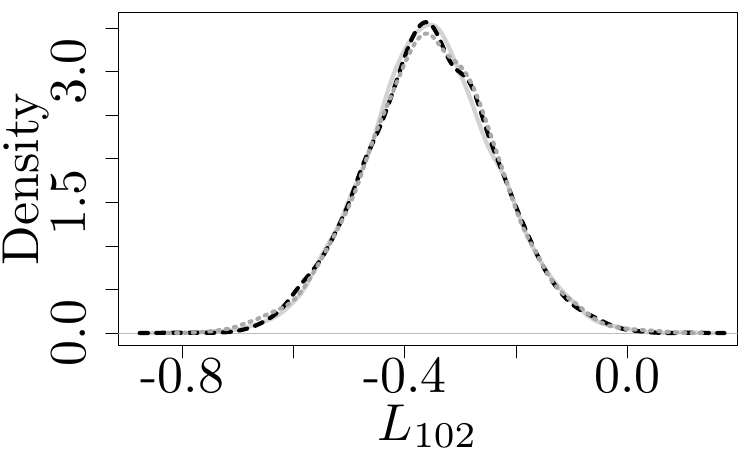}
\includegraphics[trim= {0.0cm 0.00cm 0.0cm 0.0cm}, clip,  width=0.24\columnwidth]{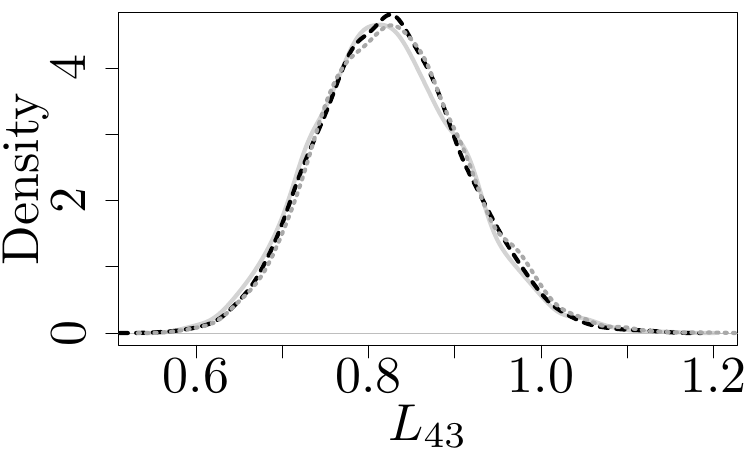}
\includegraphics[trim= {0.0cm 0.00cm 0.0cm 0.0cm}, clip,  width=0.24\columnwidth]{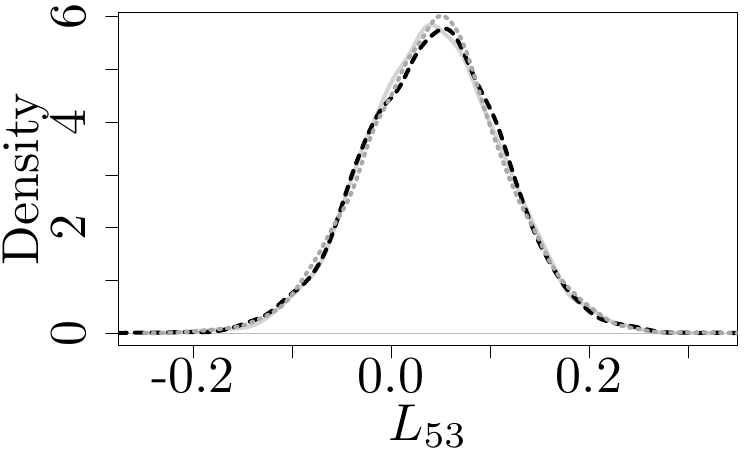}
\includegraphics[trim= {0.0cm 0.00cm 0.0cm 0.0cm}, clip,  width=0.24\columnwidth]{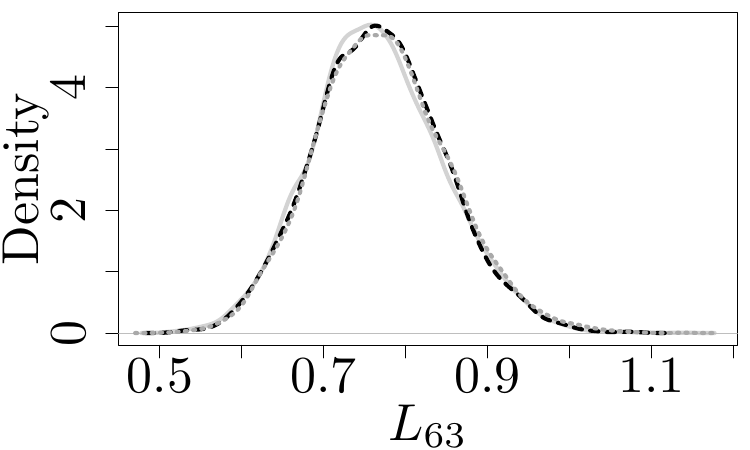}\\
\includegraphics[trim= {0.0cm 0.00cm 0.0cm 0.0cm}, clip,  width=0.24\columnwidth]{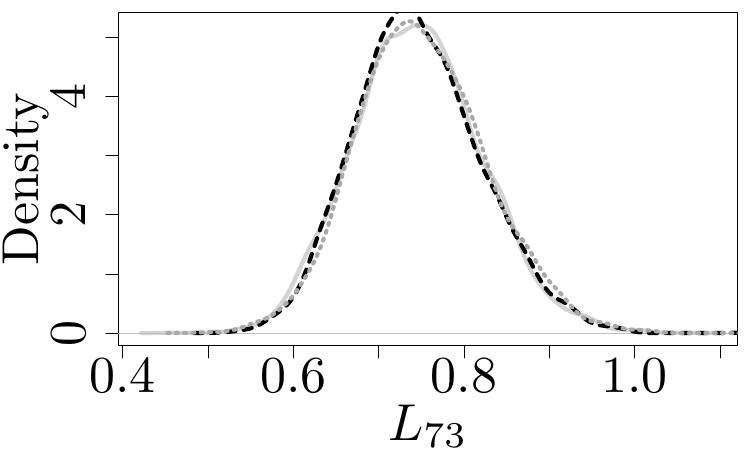}
\includegraphics[trim= {0.0cm 0.00cm 0.0cm 0.0cm}, clip,  width=0.24\columnwidth]{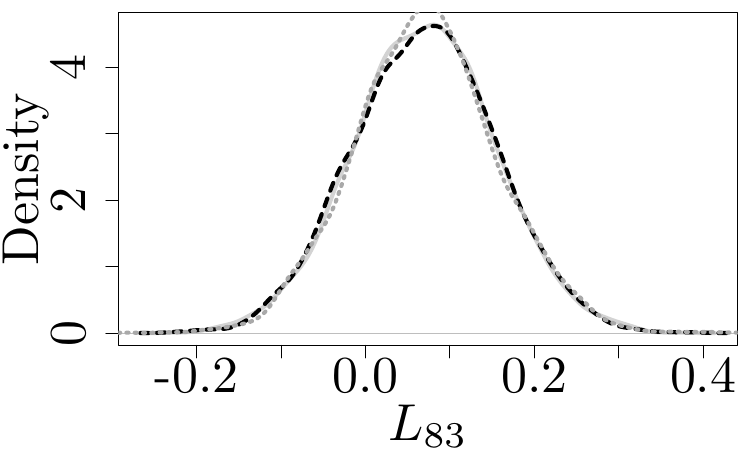}
\includegraphics[trim= {0.0cm 0.00cm 0.0cm 0.0cm}, clip,  width=0.24\columnwidth]{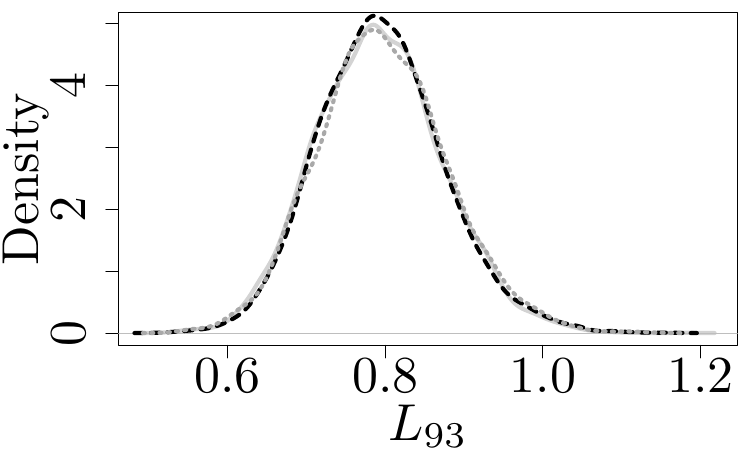}
\includegraphics[trim= {0.0cm 0.00cm 0.0cm 0.0cm}, clip,  width=0.24\columnwidth]{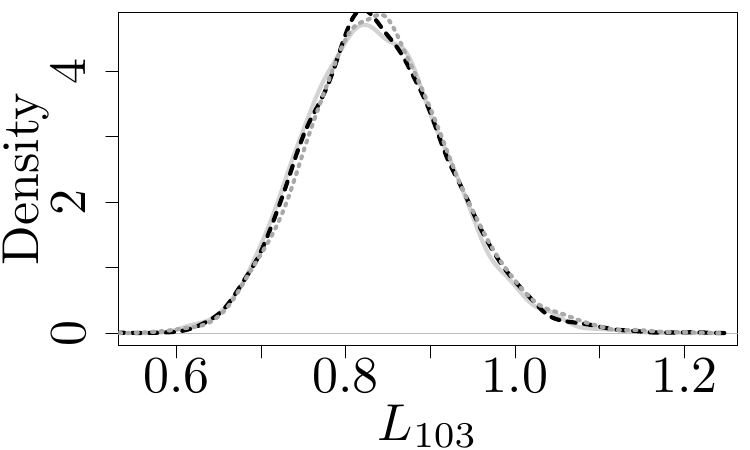}\\
\includegraphics[trim= {0.0cm 0.00cm 0.0cm 0.0cm}, clip,  width=0.24\columnwidth]{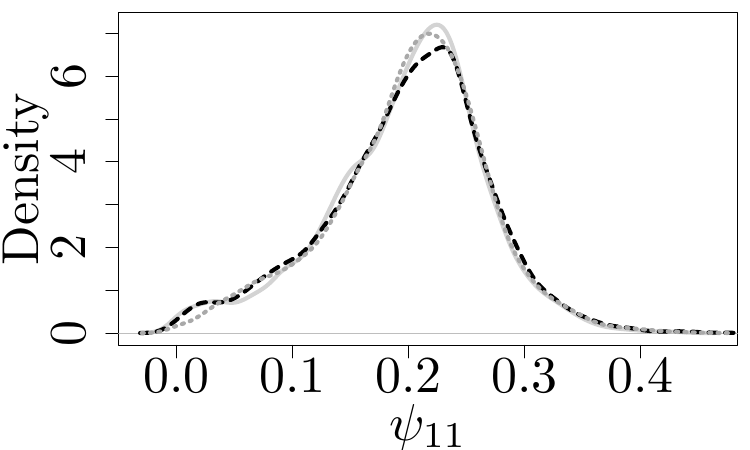}
\includegraphics[trim= {0.0cm 0.00cm 0.0cm 0.0cm}, clip,  width=0.24\columnwidth]{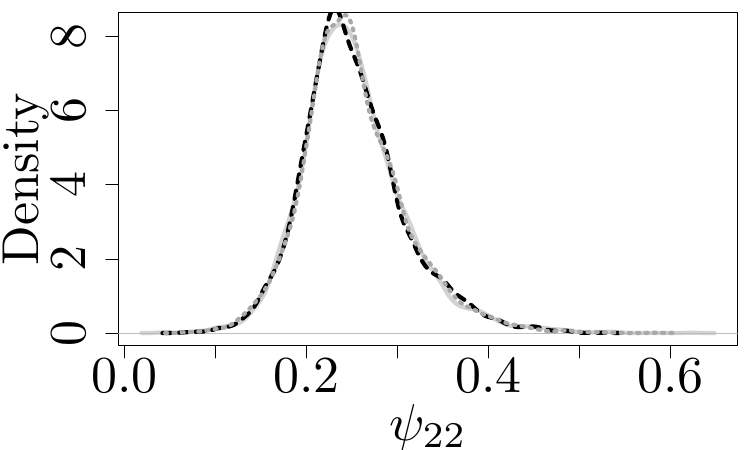}
\includegraphics[trim= {0.0cm 0.00cm 0.0cm 0.0cm}, clip,  width=0.24\columnwidth]{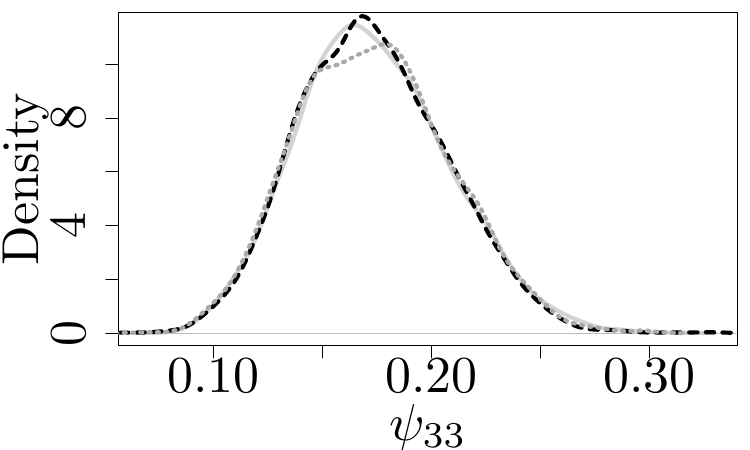}
\includegraphics[trim= {0.0cm 0.00cm 0.0cm 0.0cm}, clip,  width=0.24\columnwidth]{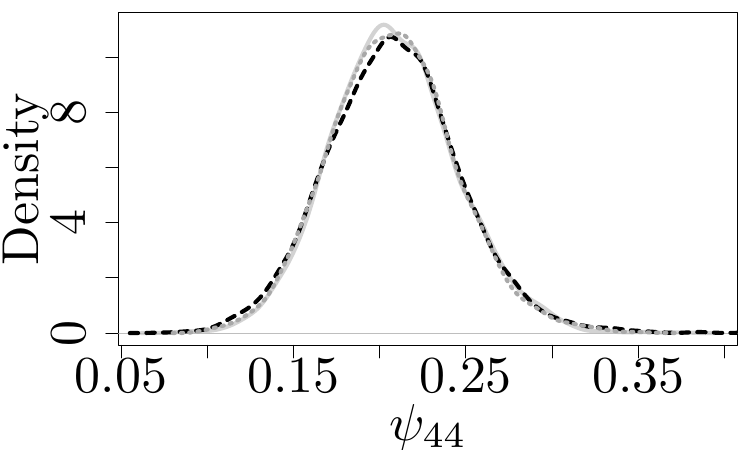}\\
\includegraphics[trim= {0.0cm 0.00cm 0.0cm 0.0cm}, clip,  width=0.24\columnwidth]{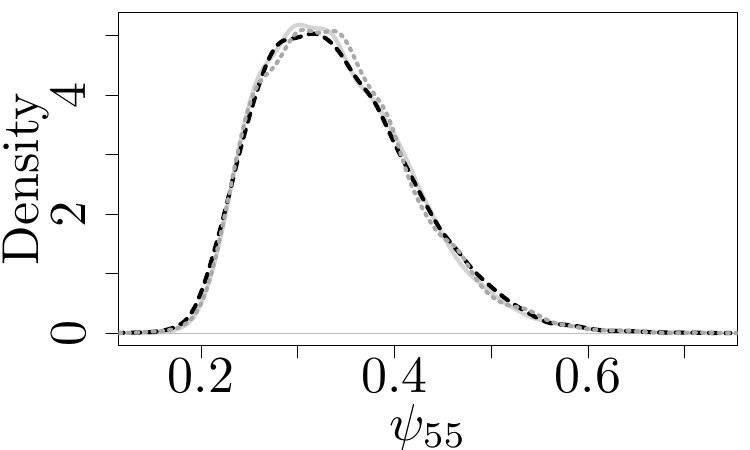}
\includegraphics[trim= {0.0cm 0.00cm 0.0cm 0.0cm}, clip,  width=0.24\columnwidth]{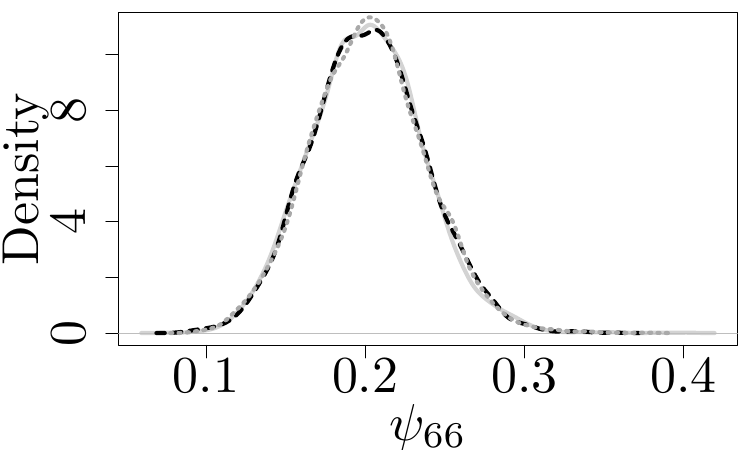}
\includegraphics[trim= {0.0cm 0.00cm 0.0cm 0.0cm}, clip,  width=0.24\columnwidth]{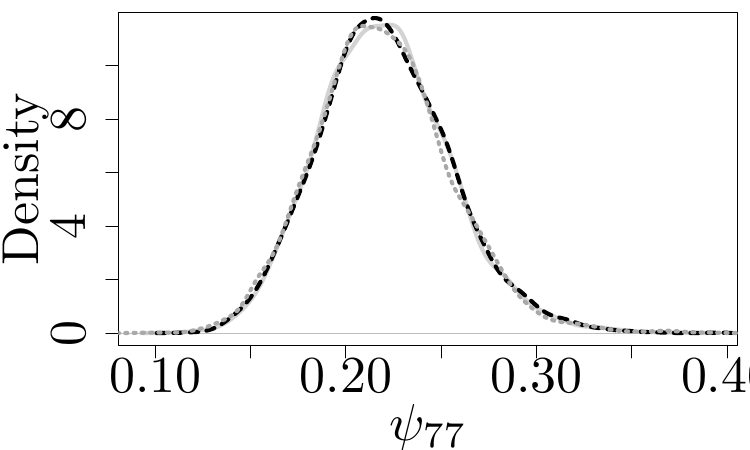}
\includegraphics[trim= {0.0cm 0.00cm 0.0cm 0.0cm}, clip,  width=0.24\columnwidth]{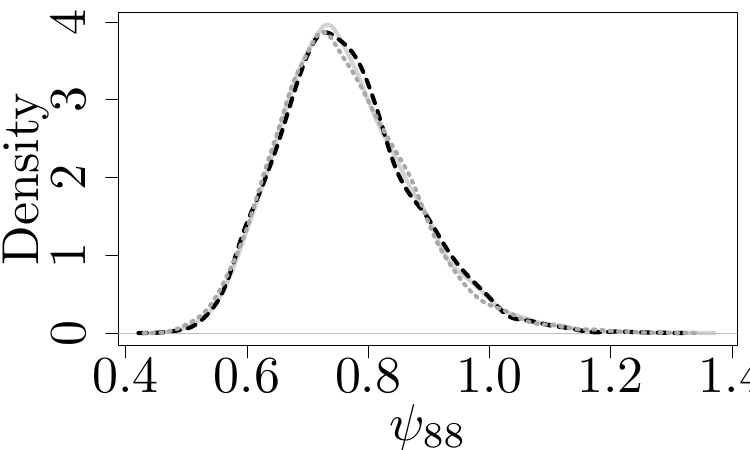}\\
\includegraphics[trim= {0.0cm 0.00cm 0.0cm 0.0cm}, clip,  width=0.24\columnwidth]{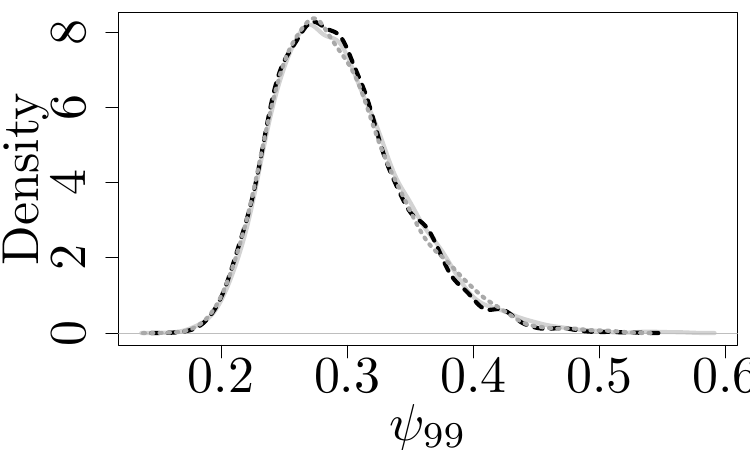}
\includegraphics[trim= {0.0cm 0.00cm 0.0cm 0.0cm}, clip,  width=0.24\columnwidth]{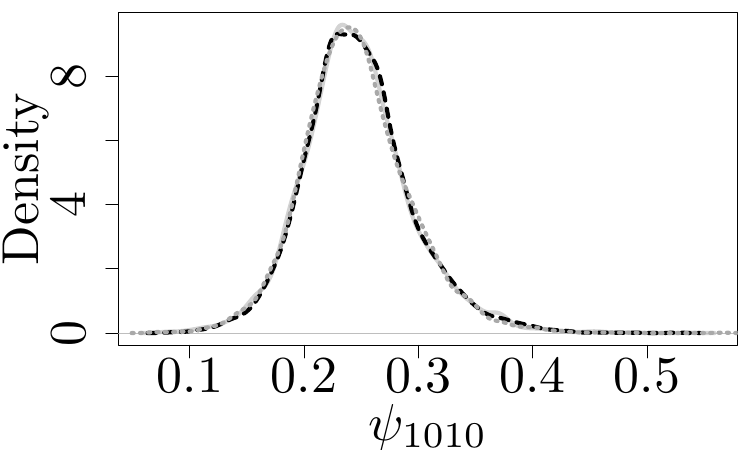}

\caption{Factor model posteriors when $n = 100$, $p = 10$ and $d = 3$ as estimated by different \texttt{Stan} implementations.}
\label{Fig:factor_posteriors}
\end{center}
\end{figure}

Figure \ref{Fig:factor_posteriors} compares the posterior samples from the vectorised, \texttt{brms}, \texttt{rstanarm} and sufficient statistics \texttt{Stan} implementations and shows that all 4 models achieve equivalent posterior approximation.

\subsection{Poisson Regression}{}

Here we provide the \texttt{Stan} code used to implement the Bayesian Poisson regression example in Section \ref{Sec:Poisson}.

Listing \ref{Lst:stan_poisson_vectorised} presents a vectorised implementation.

\begin{lstlisting}[style=myListingStyle, caption=Vectorised implementation of Bayesian Poisson regression in \texttt{Stan}., label=Lst:stan_poisson_vectorised]
data {
  int<lower=1> N;  // total number of observations
  int Y[N];  // response variable
  int<lower=1> K;  // number of population-level effects
  matrix[N, K] X;  // population-level design matrix
}

parameters {
  vector[K] b;  // regression coefficients
}

model {
  b ~ normal(0, 2);
  Y ~ poisson(exp(X*b));
}
\end{lstlisting}

Listing \ref{Lst:stan_poisson_brms} is the \texttt{Stan} code produced by \texttt{brms} for this model.

\begin{lstlisting}[style=myListingStyle, caption= \texttt{brms} implementation of Bayesian Poisson regression in \texttt{Stan}., label=Lst:stan_poisson_brms]
// generated with brms 2.22.0
functions {
}
data {
  int<lower=1> N;  // total number of observations
  array[N] int Y;  // response variable
  int<lower=1> K;  // number of population-level effects
  matrix[N, K] X;  // population-level design matrix
  int prior_only;  // should the likelihood be ignored?
}
transformed data {
}
parameters {
  vector[K] b;  // regression coefficients
}
transformed parameters {
  real lprior = 0;  // prior contributions to the log posterior
  lprior += normal_lpdf(b | 0, 2);
}
model {
  // likelihood including constants
  if (!prior_only) {
    target += poisson_log_glm_lpmf(Y | X, 0, b);
  }
  // priors including constants
  target += lprior;
}
generated quantities {
}
\end{lstlisting}

Listing \ref{Lst:stan_poisson_suffstat} presents our implementation of Bayesian Poisson regression that leverages the sufficient statistics representation of the likelihood.

\begin{lstlisting}[style=myListingStyle, caption=Bayesian Poisson regression taking advantage of sufficient statistics in \texttt{Stan}., label=Lst:stan_poisson_suffstat]
data {
  int<lower=1> N;  // total number of observations
  vector[N] Y;  // response variable
  //int Y[N];  // response variable
  int<lower=1> K;  // number of population-level effects
  matrix[N, K] X;  // population-level design matrix
}

transformed data {
  row_vector[K] Syx;

  Syx = Y'*X;
}

parameters {
  vector[K] b;  // regression coefficients
}

transformed parameters {   
}

model {
  // Priors:
  target += normal_lpdf(b | 0, 2);
  // Likelihood:
  target += Syx*b - sum(exp(X*b));
}
\end{lstlisting}

\begin{figure}[!ht]
\begin{center}
\includegraphics[trim= {0.0cm 0.00cm 0.0cm 0.0cm}, clip,  width=0.32\columnwidth]{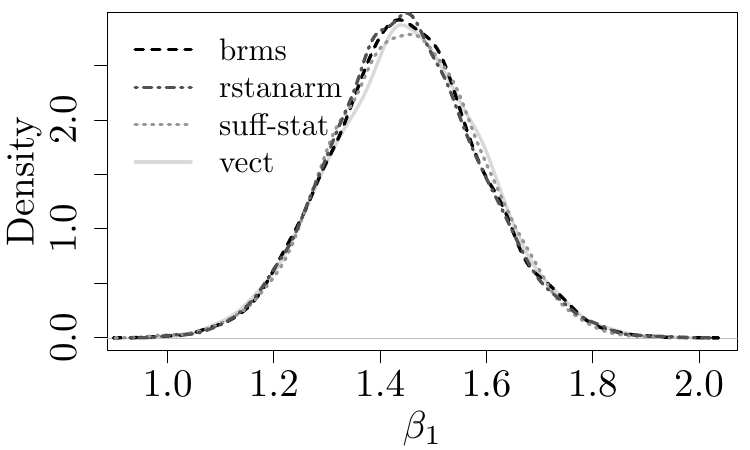}
\includegraphics[trim= {0.0cm 0.00cm 0.0cm 0.0cm}, clip,  width=0.32\columnwidth]{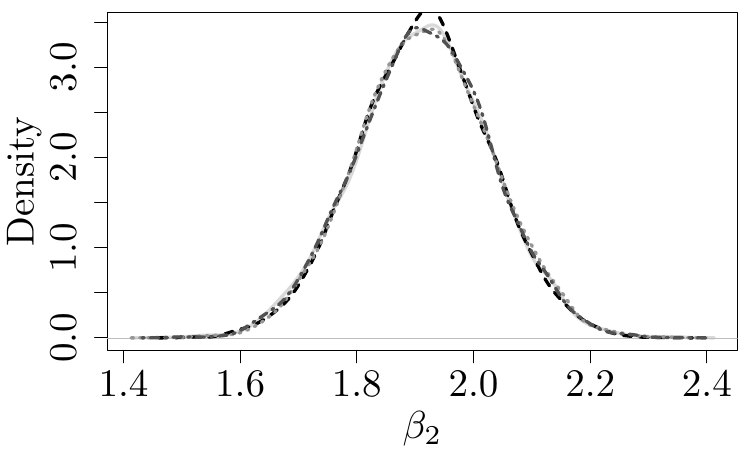}
\includegraphics[trim= {0.0cm 0.00cm 0.0cm 0.0cm}, clip,  width=0.32\columnwidth]{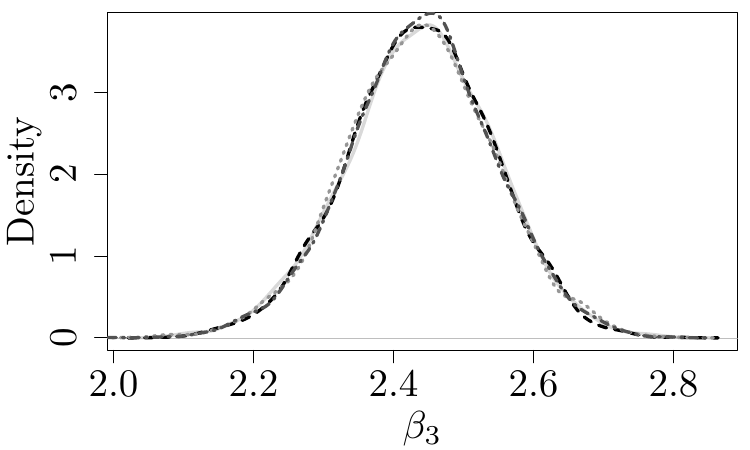}\\
\includegraphics[trim= {0.0cm 0.00cm 0.0cm 0.0cm}, clip,  width=0.32\columnwidth]{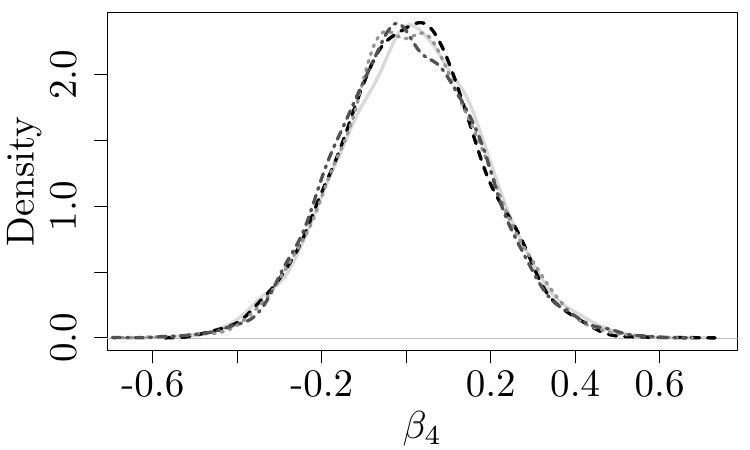}
\includegraphics[trim= {0.0cm 0.00cm 0.0cm 0.0cm}, clip,  width=0.32\columnwidth]{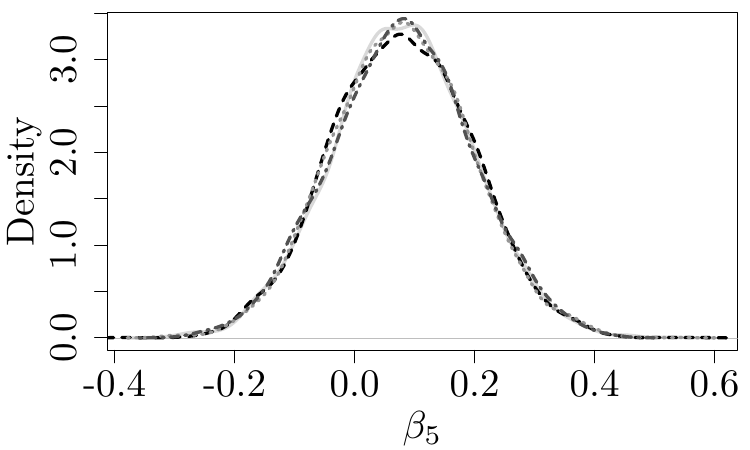}
\includegraphics[trim= {0.0cm 0.00cm 0.0cm 0.0cm}, clip,  width=0.32\columnwidth]{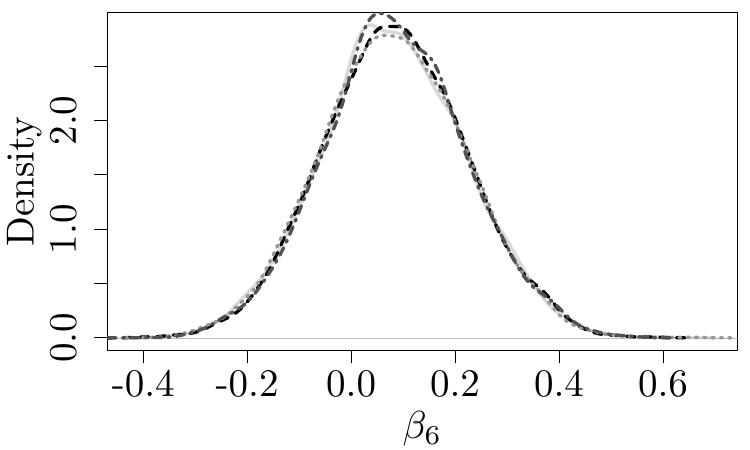}\\
\includegraphics[trim= {0.0cm 0.00cm 0.0cm 0.0cm}, clip,  width=0.32\columnwidth]{Figures/regression/BayesianLinearRegression_BRMS_SuffStat_stan_n100_p10_diag_tikz-7.pdf}
\includegraphics[trim= {0.0cm 0.00cm 0.0cm 0.0cm}, clip,  width=0.32\columnwidth]{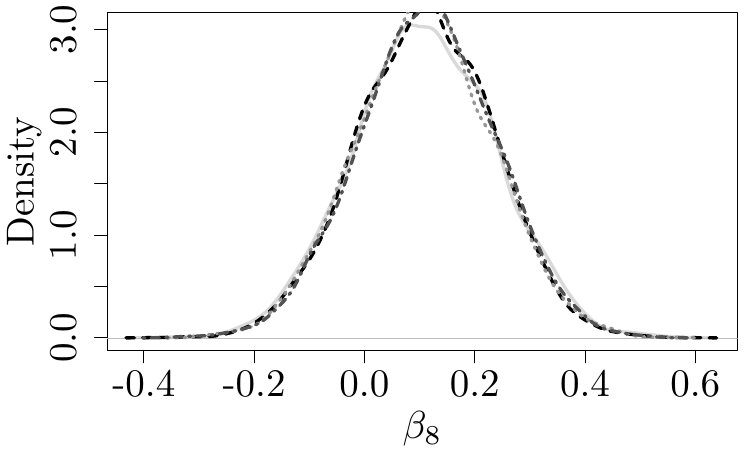}
\includegraphics[trim= {0.0cm 0.00cm 0.0cm 0.0cm}, clip,  width=0.32\columnwidth]{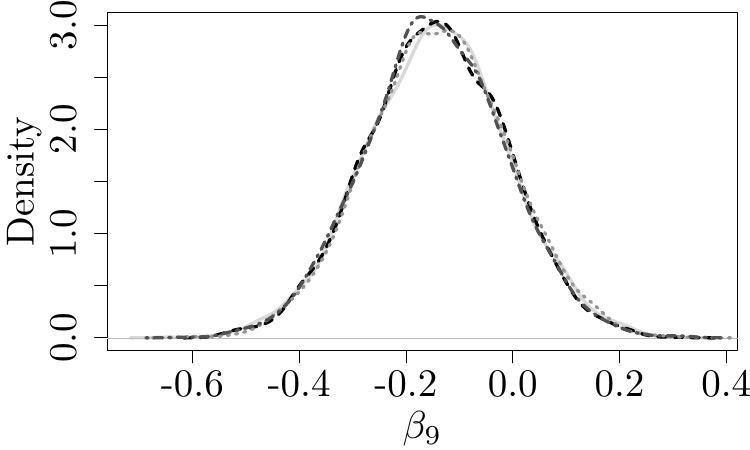}\\
\includegraphics[trim= {0.0cm 0.00cm 0.0cm 0.0cm}, clip,  width=0.32\columnwidth]{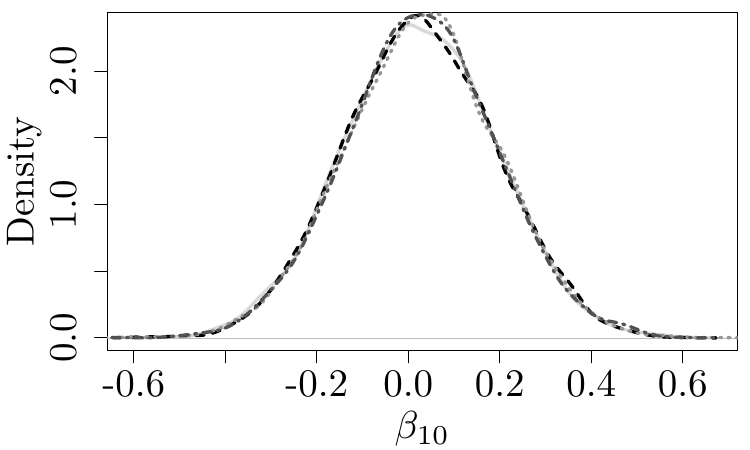}
\caption{Poisson regression posteriors with $n = 100$ and $p = 10$ as estimated by different \texttt{Stan} implementations.}
\label{Fig:poisson_posteriors}
\end{center}
\end{figure}

Figure \ref{Fig:poisson_posteriors} compares the posterior samples from the vectorised, \texttt{brms}, \texttt{rstanarm} and sufficient statistics \texttt{Stan} implementations and shows that all 4 models achieve equivalent posterior approximation.

\end{document}